\documentclass[aps,notitlepage,twocolumn,nofootinbib,superscriptaddress,longbibliography]{revtex4-1}
\usepackage{algorithm,amsfonts,amsmath,amssymb,mathtools,mathrsfs,bbm,float}
\usepackage{graphicx,epic,eepic,epsfig,latexsym,verbatim,color}
\usepackage[shortlabels]{enumitem}
 
\usepackage{booktabs} 
\usepackage{tikz}
\usetikzlibrary{chains}
\usetikzlibrary{fit}
\usepackage{epsfig}
\usetikzlibrary{shapes.symbols,patterns} % for source symbols
\usepackage{pgfplots}

\usepackage[strict]{changepage}
\usepackage{hyperref}
\hypersetup{colorlinks=true,citecolor=blue,linkcolor=blue,filecolor=blue,urlcolor=blue,breaklinks=true}

\usepackage[marginal]{footmisc}
\usepackage{url}
\usepackage{theorem}

\newtheorem{definition}{Definition}
\newtheorem{proposition}{Proposition}
\newtheorem{lemma}[proposition]{Lemma}

\def\squareforqed{\hbox{\rlap{$\sqcap$}$\sqcup$}}
\def\qed{\ifmmode\squareforqed\else{\unskip\nobreak\hfil
\penalty50\hskip1em\null\nobreak\hfil\squareforqed
\parfillskip=0pt\finalhyphendemerits=0\endgraf}\fi}
\def\endenv{\ifmmode\;\else{\unskip\nobreak\hfil
\penalty50\hskip1em\null\nobreak\hfil\;
\parfillskip=0pt\finalhyphendemerits=0\endgraf}\fi}

\newcounter{remark}
\newenvironment{remark}[1][]{\refstepcounter{remark}\par\medskip\noindent%
\textbf{Remark~\theremark #1} }{\medskip}

\newcounter{example}

\mathchardef\ordinarycolon\mathcode`\:
\mathcode`\:=\string"8000
\def\vcentcolon{\mathrel{\mathop\ordinarycolon}}
\begingroup \catcode`\:=\active
  \lowercase{\endgroup
  \let :\vcentcolon
  }

\usepackage{cleveref}
\usepackage{graphicx}
\usepackage{xcolor}

\definecolor{darkblue}{RGB}{0,76,156}
\definecolor{darkkblue}{RGB}{0,0,153}
\definecolor{blue2}{RGB}{102,178,255}
\definecolor{darkred}{RGB}{195,0,0}

\RequirePackage[framemethod=default]{mdframed}
\newmdenv[skipabove=7pt,
skipbelow=7pt,
backgroundcolor=darkblue!15,
innerleftmargin=5pt,
innerrightmargin=5pt,
innertopmargin=5pt,
leftmargin=0cm,
rightmargin=0cm,
innerbottommargin=5pt,
linewidth=1pt]{tBox}

\newmdenv[skipabove=7pt,
skipbelow=7pt,
backgroundcolor=blue2!25,
innerleftmargin=5pt,
innerrightmargin=5pt,
innertopmargin=5pt,
leftmargin=0cm,
rightmargin=0cm,
innerbottommargin=5pt,
linewidth=1pt]{dBox}

\newmdenv[skipabove=7pt,
skipbelow=7pt,
backgroundcolor=darkred!15,
innerleftmargin=5pt,
innerrightmargin=5pt,
innertopmargin=5pt,
leftmargin=0cm,
rightmargin=0cm,
innerbottommargin=5pt,
linewidth=1pt]{rBox}

\newcommand{\nc}{\newcommand}
\nc{\bra}[1]{\langle#1|}
\nc{\ket}[1]{|#1\rangle}
\nc{\ketbra}[2]{|#1\rangle\!\langle#2|}
\nc{\braket}[2]{\langle#1|#2\rangle}

\nc{\proj}[1]{| #1\rangle\!\langle #1 |}
\nc{\avg}[1]{\langle#1\rangle}
\nc{\rank}{\operatorname{Rank}}
\nc{\smfrac}[2]{\mbox{$\frac{#1}{#2}$}}
\nc{\tr}{\operatorname{Tr}}
\nc{\ox}{\otimes}
\nc{\dg}{\dagger}
\nc{\dn}{\downarrow}
\nc{\cA}{{\cal A}}
\nc{\cB}{{\cal B}}
\nc{\cC}{{\cal C}}
\nc{\cD}{{\cal D}}
\nc{\cE}{{\cal E}}
\nc{\cF}{{\cal F}}
\nc{\cG}{{\cal G}}
\nc{\cH}{{\cal H}}
\nc{\cI}{{\cal I}}
\nc{\cJ}{{\cal J}}
\nc{\cK}{{\cal K}}
\nc{\cL}{{\cal L}}
\nc{\cM}{{\cal M}}
\nc{\cN}{{\cal N}}
\nc{\cO}{{\cal O}}
\nc{\cP}{{\cal P}}
\nc{\cQ}{{\cal Q}}
\nc{\cR}{{\cal R}}
\nc{\cS}{{\cal S}}
\nc{\cT}{{\cal T}}
\nc{\cU}{{\cal U}}
\nc{\cV}{{\cal V}}
\nc{\cX}{{\cal X}}
\nc{\cY}{{\cal Y}}
\nc{\cZ}{{\cal Z}}
\nc{\cW}{{\cal W}}
\nc{\csupp}{{\operatorname{csupp}}}
\nc{\qsupp}{{\operatorname{qsupp}}}
\nc{\var}{{\operatorname{Var}}}
\nc{\rar}{\rightarrow}
\nc{\lrar}{\longrightarrow}
\nc{\polylog}{{\operatorname{polylog}}}
\nc{\wt}{{\operatorname{wt}}}
\nc{\supp}{{\operatorname{supp}}}

\nc{\argmin}{{\operatorname{argmin}}}

\def\x{\xi}

\def\r{\rho}

\nc{\RR}{{{\mathbb R}}}
\nc{\CC}{{{\mathbb C}}}
\nc{\FF}{{{\mathbb F}}}
\nc{\NN}{{{\mathbb N}}}
\nc{\ZZ}{{{\mathbb Z}}}
\nc{\PP}{{{\mathbb P}}}
\nc{\QQ}{{{\mathbb Q}}}
\nc{\UU}{{{\mathbb U}}}
\nc{\EE}{{{\mathbb E}}}
\nc{\id}{{\operatorname{id}}}

\nc{\CHSH}{{\operatorname{CHSH}}}

\newcommand{\Op}{\operatorname}
\nc{\rU}{\mbox{U}}

\nc{\ob}[1]{#1}

\nc{\SEP}{{\text{\rm SEP}}}
\nc{\NS}{{\text{\rm NS}}}
\nc{\LOCC}{{\text{\rm LOCC}}}
\nc{\PPT}{{\text{\rm PPT}}}
\nc{\EXT}{{\text{\rm EXT}}}
\nc{\Sym}{{\operatorname{Sym}}}

\nc{\ERLO}{{E_{\text{r,LO}}}}
\nc{\ERLOCC}{{E_{\text{r,LOCC}}}}
\nc{\ERPPT}{{E_{\text{r,PPT}}}}
\nc{\ERLOCCinfty}{{E^{\infty}_{\text{r,LOCC}}}}
\nc{\Aram}{{\operatorname{\sf A}}}

\usepackage{hyperref}
\hypersetup{colorlinks=true,citecolor=blue,linkcolor=blue,filecolor=blue,urlcolor=blue,breaklinks=true}

\makeatletter
\def\grd@save@target#1{%
  \def\grd@target{#1}}
\def\grd@save@start#1{%
  \def\grd@start{#1}}
\tikzset{
  grid with coordinates/.style={
    to path={%
      \pgfextra{%
        \edef\grd@@target{(\tikztotarget)}%
        \tikz@scan@one@point\grd@save@target\grd@@target\relax
        \edef\grd@@start{(\tikztostart)}%
        \tikz@scan@one@point\grd@save@start\grd@@start\relax
        \draw[minor help lines,magenta] (\tikztostart) grid (\tikztotarget);
        \draw[major help lines] (\tikztostart) grid (\tikztotarget);
        \grd@start
        \pgfmathsetmacro{\grd@xa}{\the\pgf@x/1cm}
        \pgfmathsetmacro{\grd@ya}{\the\pgf@y/1cm}
        \grd@target
        \pgfmathsetmacro{\grd@xb}{\the\pgf@x/1cm}
        \pgfmathsetmacro{\grd@yb}{\the\pgf@y/1cm}
        \pgfmathsetmacro{\grd@xc}{\grd@xa + \pgfkeysvalueof{/tikz/grid with coordinates/major step}}
        \pgfmathsetmacro{\grd@yc}{\grd@ya + \pgfkeysvalueof{/tikz/grid with coordinates/major step}}
        \foreach \x in {\grd@xa,\grd@xc,...,\grd@xb}
        \node[anchor=north] at (\x,\grd@ya) {\pgfmathprintnumber{\x}};
        \foreach \y in {\grd@ya,\grd@yc,...,\grd@yb}
        \node[anchor=east] at (\grd@xa,\y) {\pgfmathprintnumber{\y}};
      }
    }
  },
  minor help lines/.style={
    help lines,
    step=\pgfkeysvalueof{/tikz/grid with coordinates/minor step}
  },
  major help lines/.style={
    help lines,
    line width=\pgfkeysvalueof{/tikz/grid with coordinates/major line width},
    step=\pgfkeysvalueof{/tikz/grid with coordinates/major step}
  },
  grid with coordinates/.cd,
  minor step/.initial=.2,
  major step/.initial=1,
  major line width/.initial=2pt,
}
\makeatother

\usepackage{thmtools}
\usepackage{thm-restate}
\usepackage{etoolbox}
\makeatletter
\def\problem@s{}
\newcounter{problems@cnt}

\newcommand{\allproblems}{\problem@s}
\makeatother
\usepackage{amsmath,amsfonts,amssymb,array,graphicx,mathtools,multirow,bm,times,tcolorbox,relsize}
\usepackage{algorithmic}
\usepackage{dcolumn}
\usepackage[utf8]{inputenc}
\usepackage[T1]{fontenc}
\usepackage[qm]{qcircuit}

\usepackage{tikz-cd}
\tikzcdset{every label/.append style = {font = \small}}
\usepackage{tikz}
\usetikzlibrary{positioning}
\usetikzlibrary{shapes.geometric}
\usetikzlibrary{calc}
\definecolor{tensorblue}{rgb}{0.8,0.9,1}
\tikzset{ten/.style={fill=tensorblue}}
\newcommand{\diagram}[1]{ \begin{array}{cc}\begin{tikzpicture}[scale=.5,every node/.style={sloped,allow upside down},baseline={([yshift=+0ex]current bounding box.center)}] #1 \end{tikzpicture} \end{array} }
\allowdisplaybreaks

\begin{document}

\title{Quantum sequential scattering model for quantum state learning
}
\author{Mingrui Jing}
\thanks{These two authors contributed equally.}
\affiliation{Thrust of Artificial Intelligence, Information Hub, Hong Kong University of Science and Technology (Guangzhou), Nansha, China}
\affiliation{Institute for Quantum Computing, Baidu Research, Beijing 100193, China}
\author{Geng Liu}
\thanks{These two authors contributed equally.}
\affiliation{Thrust of Artificial Intelligence, Information Hub, Hong Kong University of Science and Technology (Guangzhou), Nansha, China}
\affiliation{Institute for Quantum Computing, Baidu Research, Beijing 100193, China}
\thanks{These two authors contributed equally.}
\author{Hongbin Ren}
\affiliation{Institute for Quantum Computing, Baidu Research, Beijing 100193, China}
\author{Xin Wang}
\email{felixxinwang@hkust-gz.edu.cn}
\affiliation{Thrust of Artificial Intelligence, Information Hub, Hong Kong University of Science and Technology (Guangzhou), Nansha, China}
\affiliation{Institute for Quantum Computing, Baidu Research, Beijing 100193, China}
\begin{abstract}
Learning probability distribution is an essential framework in classical learning theory. As a counterpart, quantum state learning has spurred the exploration of quantum machine learning theory. However, as dimensionality increases, learning a high-dimensional unknown quantum state via conventional quantum neural network approaches remains challenging due to trainability issues. In this work, we devise the quantum sequential scattering model (QSSM), inspired by the classical diffusion model, to overcome this scalability issue. Training of our model could effectively circumvent the vanishing gradient problem to a large class of high-dimensional target states possessing polynomial-scaled Schmidt ranks. Theoretical analysis and numerical experiments provide evidence for our model's effectiveness in learning both physical and algorithmic meaningful quantum states and show an out-performance beating the conventional approaches in training speed and learning accuracy. Our work has indicated that an increasing entanglement, a property of quantum states, in the target states, necessitates a larger scaled model, which could reduce our model's learning performance and efficiency.
\end{abstract}

\date{\today}
\maketitle

% \tableofcontents
%%%%%%%%%%%%%%%%%%%%%%%%%%%%%%%%%%%%%%%%%%%%%%%%%%%%%%%%%%%%%%%%%%%%%%%%%%%
\section{Introduction}
The innovation of classical machine learning has brought significant convenience and efficiency in industry and society. In particular, learning distributions between individual events and data is one of the crucial tasks for multiple usages in decades~\cite{anderson1977distinctive,geng2016label}. A plethora of approaches and schemes have been designed to learn probability distributions, such as continuous evolutionary algorithms~\cite{Hansen2015,Kern2004} and supervised learning within the neural network framework including Boltzmann machine, graph neural network and diffusion model~\cite{baum1987supervised,franceschi2019learning,hoogeboom2021argmax}

Meanwhile, by the fast growth of the requirement on computational power, quantum computing, as a prospective new framework, is expected to provide advantages over classical technology. The remarkable achievements from classical machine learning models~\cite{lecun2015deep, serban2016building} have spurred the generation of their counterparts within the field of quantum machine learning (QML). See Refs. \cite{biamonte2017quantum,schuld2015introduction,lloyd2013quantum,schuld2014quest,Cerezo2022a,Abbas2021,Du2023,Yu2022_power,chowdhury2020variational,ghosh2019quantum,Wang2021} for reviews and recent progresses. Quantum neural networks (QNNs) composed of layers of parametrised quantum circuits have received massive attention regarding various architectures addressing computation challenges~\cite{rebentrost2018quantum,zhao2019building,cong2019quantum}, including quantum state learning. 

In quantum, the correlations between quantum data are encoded in the quantum states. Consequently, the task of learning an arbitrary quantum state bears a resemblance to classical distribution learning, which has inspired developments of state learning QML models~\cite{chowdhury2020variational,ghosh2019quantum,Wang2021}. As a main solution to quantum state learning, however, the implementation of the QNN-based methods suffers obstacles in efficiency, scalability and trainability. Specifically, training deep QNNs composed of multiple layers can experience exponentially vanishing gradients, or called \textit{barren plateaus} (BP)~\cite{McClean2018} when targeting high-dimensional states. 

This work proposed a quantum sequential scattering model (QSSM) to overcome this bottleneck in QNN-powered state learning techniques. We provide both theoretical and numerical demonstrations of QSSM on training efficiency and learning accuracy, which can outperform the conventional QNN model using universal layers. Recent research on the trainability issue of QNNs indicates prospective directions by reducing the expressibility of QNN architectures~\cite{Cerezo2021, Liu2022}, adopting clever parameterization strategies~\cite{Grant2019, kulshrestha2022beinit,volkoff2021large,friedrich2022avoiding} and using adaptive algorithms~\cite{Grimsley2019,Zhang2021,Skolik2021,Grimsley2022}. 

We drew inspiration from the classical diffusion model~\cite{yang2022diffusion} by conducting the state learning with progressively augmenting sublevels in a sequential manner. Our model combines the ideas of quantum purification theory and adaptive and layerwise training~\cite{quek2021adaptive,Skolik2021} for which the training process can be treated as the dilation of quantum information from subsystems to the entire one. The structure of the model ensures a dramatic reduction in the number of optimized parameters at each training step and, therefore, avoids barren plateaus for a large class of target states.

Our work is presented in the following order: We first introduce the basic notations and definitions in Section~\ref{sec: preliminary}. In Section~\ref{sec: main result}, we present the theoretical guarantee of the QSSM in view of information diffusion and trainability. Then, we explicitly describe our QSSM processing state learning task~\ref{sec:quantum_sequential_scattering_model}, including the algorithm optimization and gradient estimations. In Section~\ref{sec:numerical_exp}, we illustrate the numerical simulations on the effectiveness and trainability of the model learning both physical and algorithmic meaningful states. Noisy simulations are also provided. Conclusion and outlook will be given in Section~\ref{sec: conclusion and future works}.

\section{Preliminaries}\label{sec: preliminary}
\subsection{Classical Distribution learning} 
We briefly introduce the formalism concerning classical probability distribution learning. Correlations between discrete data variables, denoted as $X$, can be characterized by some probability distributions
$D$~\cite{Kearns1994}. The learning of such a distribution can be described as constructing a generator $G_{D'}$ that takes $x\in X$ as an argument and outputs $G_{D'}[x]\in X$ with respect to a distribution $D'$. The generator can be realized via a classical machine learning model, which is trained to achieve $d(D, D')\leq \varepsilon$ for some legal metric $d$, e.g., \textit{Kullback-Leibler divergence}~\cite{Csiszar1975}, and a threshold error $\varepsilon$.  

\subsection{Quantum State Learning}
A typical quantum state learning task for an unexplored target state $\rho$, as a density matrix, solves for a generator that can be efficiently constructed to produce a representation $\rho'$ which $\cD(\rho, \rho')\leq \varepsilon$ resembling classical distribution learning.  Here $\cD$ is a feasible distance measure on matrix space. Such a generator can veritably produce $\rho'$ instead of numerically simulating it~\cite{Vidal2003} and can be repeatedly used in further computational tasks. This work focuses on the QNN-powered algorithms combining both classical and quantum computation. Utilizing parameterized quantum circuits working as the state generators that are trained by gradient descent or gradient-free methods to determine the optimal parameters~\cite{Peruzzo2014,kandala2017hardware}. Beyond our scope, schemes using shadow tomography~\cite{aaronson2018shadow,huang2022learning} fulfil another category of state learning with the aim of characterizing the classical information of quantum states.

\subsection{Quantum Computing \& QNN layers}
Quantum information is encoded and processed via the fundamental cells, namely, qubits. An $n$-qubit state can be mathematically represented by a $2^n \times 2^n$ positive semi-definite density matrix $\rho$, i.e., $\rho\succeq 0$ over the complex field and $\tr[\rho] = 1$. A pure state, in this formulation, satisfy $\rank{(\rho)} = 1$ and can be expressed in  \textit{Dirac bra-ket} notation as $\rho = \ketbra{\psi}{\psi}$ where $\ket{\psi}\in \CC^{2^n}$ denotes a \textit{Hilbert space} unit column vector with the corresponding \textit{dual vector} $\bra{\psi}^\dagger = \ket{\psi}$ and $\dagger$ denoting the complex conjugate transpose operation. A mixed state satisfies $\rank{(\rho)}>1$, and based on \textit{Spectral theorem}, it has a decomposition form $\rho = \sum_j p_j \ketbra{\psi_j}{\psi_j}$ where $p_j > 0$ denotes the probability of observing $\ketbra{\psi_j}{\psi_j}$ in $\rho$ and $\sum_j p_j = 1$. 

\begin{figure}
    \centering
    $$\Qcircuit @C=.9em @R=.4em{
    & & & & & & & & & &\times d\\
    &\qw &\gate{U3(\bm{\theta}_1)} &\ctrl{1} &\qw  &\gate{U3(\bm{\theta}_5)} &\ctrl{1} &\qw &\gate{U3(\bm{\theta}_9)} &\qw &\qw\\ 
    &\qw &\gate{U3(\bm{\theta}_2)} &\targ &\ctrl{1} &\gate{U3(\bm{\theta}_6)} &\targ &\ctrl{1} &\gate{U3(\bm{\theta}_{10})} &\qw &\qw\\ 
    &\qw &\gate{U3(\bm{\theta}_3)} &\ctrl{1} &\targ &\gate{U3(\bm{\theta}_7)} &\ctrl{1} &\targ &\gate{U3(\bm{\theta}_{11})} &\qw &\qw\\ 
    &\qw &\gate{U3(\bm{\theta}_4)} &\targ &\qw &\gate{U3(\bm{\theta}_8)} &\targ &\qw &\gate{U3(\bm{\theta}_{12})} &\qw &\qw \gategroup{2}{4}{5}{9}{1.3em}{--}\\
    }
    % \inputgroupv{1}{4}{0.4em}{3em}{n=4 \quad}
    $$
    \caption{The general architecture of the QNN layers used for quantum state learning. The U3-gates can be decomposed as a combination of $R_Z(\phi_1)R_X(-\pi/2)R_Z(\theta_1)R_X(\pi/2)R_Z(\lambda_1)$ where the parameter vector $\bm{\theta}_1 = (\theta_1, \phi_1, \lambda_1)$. The layer consists of CNOT gates and U3 gates. The dashed block circuit repeats $d$ times as the depth of the layer. The above has a layer width $w = 4$, which applies to $4$ quantum registers. In reality, the above circuit diagram represents a way of applying quantum gates sequentially in order from left to right.}
    \label{fig: layer architecture}
\end{figure}
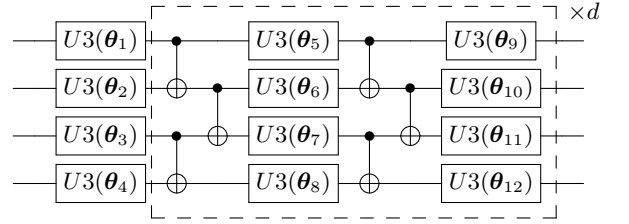

The evolution of a quantum state $\rho$ is realized by applying a series of quantum gates which are mathematically described as unitary operators. The state $\rho'$ that undergoes transformation via a quantum gate $U$ can be obtained through direct matrix multiplication, expressed as $\rho' = U\rho U^\dagger$.
Common single-qubit gates include the Pauli rotations $\{R_P(\theta) = e^{-i\frac{\theta}{2}P} |P \in \{X,Y,Z\}\}$, which are in the matrix exponential form of Pauli matrices
\begin{equation}
    X := 
    \begin{pmatrix}
        0 & 1\\
        1 & 0
    \end{pmatrix},
    Y := 
    \begin{pmatrix}
        0 & -i\\
        i & 0
    \end{pmatrix},
    Z :=
    \begin{pmatrix}
        1 & 0\\
        0 & -1
    \end{pmatrix}.
\end{equation}
Multi-qubit gates, e.g., controlled-$X$ gate CX (or CNOT) $=I\oplus X$ and controlled-$Z$ gate CZ$=I\oplus Z$ where `$\oplus$' denotes the direct sum operation live in high-dimensional linear operator space over $\CC$. Quantum measurements working as projections are applied at the end of the quantum circuits. Quantum neural networks are usually formed by layers of parameterized circuits shown in Fig.~\ref{fig: layer architecture} consisting of a bunch of single-qubit gates and several two-qubit gates.

% \MR{Theory then algorithm}
\section{Main Results}\label{sec: main result}
In this paper, we design a
quantum sequential scattering model (QSSM) absorbing the ideas of classical diffusion model and adaptive learning~\cite{quek2021adaptive}, which has modular structured parametrised circuits, or we called the \textit{scattering layer}, at each training step. Each layer ensures learning the reduced density matrix of a specific part in the target state so that the model can gradually rebuild the entire state after accomplishing all training steps. 

Our main contributions involve \textbf{(1)} conceptually proposing the idea of combining quantum information diffusion and adaptive quantum state learning, \textbf{(2)} technically devising a new quantum neural network model, namely QSSM and the state learning algorithm via a sequentially subsystem-learning strategy, \textbf{(3)} theoretically proving the effectiveness of the state learning algorithm and a polynomial-scaled gradient variance of QSSM which indicates an avoidance of barren plateaus for rank-restricted state learning, \textbf{(4)} numerically demonstrating our results on learning different quantum states involving the noise effects. We compare QSSM directly to the conventional QNN model for handling state learning tasks and showcase its enhancement in both training efficiency and learning accuracy. The main results are presented in the following sections.

\subsection{Quantum Sequential State Compositing}

Quantum states are represented in a multiple-qubit system with a fixed order. 
We treat each qubit as a quantum register, just like classical bit and classical register, and label it $q_k$ for the $k$-th register. We then define a special characteristic for quantum states.

\begin{definition}\label{def:rank sequence}
    Given an $n$-qubit quantum state $\rho$ represented by $n$ ordered quantum registers labeled as $q_1, q_2, \cdots, q_n$, denoting $\rho_k$ as the $k$-th reduced density matrix of the first $k$-register state, i.e., $\rho_k = \tr_{q_{k+1} : q_{n}}[\rho]$ for $1\leq k\leq n$ where the operation $\tr_{q_i:q_j}[\cdot]$ representing a partial tracing over registers $q_i$ to $q_j$, the (Schmidt) rank sequence of $\rho$ is an ordered list $\cR_\rho$,
    \begin{equation}
        \cR_\rho = \{r_1, r_2, \cdots, r_{n-1}, r_{n}\},
    \end{equation}
    where $r_k$ indicates  $\rank[\rho_k]$. In particular, if $\rho$ is pure, then $r_n = 1$ since $\rho$ can be represented as $\ketbra{\phi}{\phi}$ for some pure state vector $\ket{\phi}$.
\end{definition}
With these clarified, we could then present our sufficient and necessary conditions for QSSM to completely learn a target state using  Algorithm~\ref{alg:algorithm}, provided enough training time and layer width. Our analysis will concentrate on the pure target state $\rho$. However, the statement applies to the cases of mixed target states i.e., $\rank[\rho] > 1$, since we could equivalently learn its purification state by introducing auxiliary systems.
The formal version of Proposition~\ref{proposition: Truncation} can be found in Appendix~\ref{appendix:effectiveness}. 
\begin{proposition}\label{proposition: Truncation}
For a given $n$-qubit pure target state $\rho$ represented by $n$ ordered quantum registers $q_1, q_2, \cdots, q_n$, if the rank sequence of $\rho$ is $\cR_\rho = \{r_1, r_2, \cdots r_{n-1},r_n\}$. Then there exists a quantum algorithm~\ref{alg:algorithm}, based on QSSM, that could produce a state $\sigma$ exactly satisfying $\sigma = \rho$, if and only if the $k$-th scattering layer $U_k(\bm{\theta}_k)$ of QSSM has a width $w_k$ scales $\cO(\lceil \log_2 r_k\rceil)$.
\end{proposition}

We see that the width of each scattering layer scales only logarithmic regarding the target states' rank sequence. In general, even the rank of quantum pure state scales $\cO(2^{\lceil n/2\rceil})$, the logarithmic scaling in $w_k$ still guarantees a linear growth in the requirement of layer width concerning the number of qubits $n$, in the worst case. 

Moreover, though many quantum states have full rank, there is a polynomial number of dominant components in their spectral decomposition. Learning their low-rank approximation pre-determined by the quantum principal component analysis (QPCA)~\cite{lloyd2014quantum} can be treated as a quantum compressing of unknown states, which still captures the main statistical behaviours of target states. With a certain error tolerance for the low-rank approximation, the layer width can be further reduced, leading to more advantages in QSSM state learning. In the Numerical Simulations (Section~\ref{sec:numerical_exp}), we provide evidence of learning different states' rank-restricted approximation.

Compared to the $n$-qubit universal-QNN model state learning, QSSM demands significantly fewer parametric degrees of freedom (DOF) to reach the same approximating error. The generating Lie algebra of an $n$-qubit universal QNN model has to span $\mathrm{SU}(2^n)$, resulting in a model DOF of $\mathcal{O}(4^n)$. On the contrary, since the $k$-th scattering layer involves at most $(\lfloor\frac{n}{2}\rfloor + 1)$ quantum registers, the total DOF of QSSM experiences a quadratic reduction to at most $\mathcal{O}(4^{\lfloor\frac{n}{2}\rfloor})$. Also, to learn the polynomial rank-bounded target state $\rho$, i.e., $r_{\max} = \max\cR_{\rho} \sim \cO(\operatorname{Poly}(n))$. The DOF required for each scattering layer in QSSM scales $\cO(\operatorname{Poly}(n))$. Therefore, the entire model comprises fewer quantum gates, rendering this approach considerably more hardware-efficient. 

\subsection{Avoiding  Barren Plateaus}\label{sec: Trainability of QSSM}
Trainability is a critical challenge for the usage of quantum neural networks. Using a global deep QNN model brings stronger expressibility despite significantly increasing the randomness of initialization. Therefore, the initial gradient of trainable parameters in the model would exponentially vanish as the system scales up, called the Barren Plateau (BP) issue~\cite{McClean2018}.
% In this case, the partial derivative of the cost function would have a zero mean and an exponential small variance with respect to the number of qubits, thereby making it challenging to identify the correct direction to decrease the cost function value.

With the diffusion of local quantum state information, QSSM has illustrated a potential to address trainability issues by focusing on subsystems in each scattering layer instead of the whole state. From the perspective of adaptive learning, we align the reduced quantum states of the $k$-th subsystem by minimizing the $k$-th adaptive cost function of ~\eqref{Eq: cost_func} during the respective layer training,

\begin{equation}\label{Eq: cost_func}
\begin{aligned}
    C_k(\bm{\theta}) &= \|{\sigma}_k(\bm{\theta}) - \rho_k\|^2_2\\ &= \tr\left[\left({\sigma}_k(\bm{\theta}) - \rho_k\right)\left({\sigma}_k(\bm{\theta}) - \rho_k\right)^\dag\right],
\end{aligned}
\end{equation}
where $\|A\|_2$ for some linear operator $A$ denotes the Schatten-2 norm, ${\sigma}_k(\bm{\theta}_k)$ and $\rho_k$ represent the $k$-th scattering layer produced state and the $k$-th reduced target state, respectively.

In this section, we show that QSSM has explicit advantages in trainability by investigating the statistical properties of the partial gradient with respect to particular layer parameters. For the cost gradient $\partial_{\mu} C_k$ regarding the $\mu$-th trainable parameter in the $k$-th scattering layer denoted as $U_k(\bm{\theta}) = U^{(k)}_+(\bm{\theta}_+)e^{-i\theta_{\mu} H_{\mu}}U^{(k)}_-(\bm{\theta}_-)$, all the parameters in the layer are represented in a parameter vector $\bm{\theta} = (\bm{\theta}_+, \theta_\mu, \bm{\theta}_-)$, where $\bm{\theta}_-$ and $\bm{\theta}_+$ represent the parameters of the forward and the backward parts within the $k$-th scattering layer having $e^{-i\theta_\mu H_\mu}$ centralized. The results are summarized.

\begin{proposition}\label{prop:main result on gradient of QSSM}
Given the state learning algorithm stated in Proposition~\ref{proposition: Truncation}, for an $n$-qubit pure target state $\rho$ represented by $n$ ordered quantum registers $q_1, q_2, \cdots, q_n$ with a rank sequence $\cR_\rho = \{r_1, r_2, \cdots r_{n-1},r_n\}$, if one of the $U_{\pm}^{(k)}$ in the $k$-th scattering layer $U_k$ forms at least local unitary $4$-design, the expectation and the variance of $C_k$ with respect to $\theta_\mu$ can be upper bounded by,
\begin{equation}
    \EE[\partial_{\mu} C_k] = 0; \quad
    \var[\partial_{\mu} C_k] \in \cO\left(\frac{g(\rho_k)}{r_k}\right),
\end{equation}
where the expectation is computed regarding the Haar measure and the factor $g(\rho_k)$ scales polynomially in $\tr[\rho_k^2]$ known as the purity of $\rho_k$.
\end{proposition}

The formal statement of Proposition~\ref{prop:main result on gradient of QSSM} is presented in Appendix~\ref{appendix:trainability}. This proposition notably implies that the gradient magnitude is significantly determined by $r_{\max}$ in $\cR_{\rho}$ rather than the total number of quantum registers $n$. In other words, the gradient magnitude
can escape from barren plateaus by carefully setting the width of each scattering layer to adapt to the target state. A typical example is to learn an $n$-qubit GHZ state, which, by its symmetry, requires setting $w_k \leq 2$ for all scattering layers in QSSM and hence achieves $\cO(1)$ upper bound in the variance of the gradient. 

Moreover, Proposition~\ref{prop:main result on gradient of QSSM} implies that QSSM can efficiently facilitate the learning of any pure states with polynomial-scaling $r_{\max}$ in $n$. This encompasses a broad class of quantum states, including  \textit{slightly entangled states}~\cite{Vidal2003} and \textit{matrix product states}~\cite{perez2006matrix}, which extends the efficient-learnable region of quantum states using quantum neural network models. Even in the case where $r_{\max}$ scales exponentially, the gradient magnitude still gains a square root enhancement by the bounded variance of $\cO(2^{-\lfloor n/2\rfloor})$ compared with the conventional model, scaling as $\cO(2^{-n})$ to reach the same learning accuracy. 

One may also apply the previous statement by allowing the error tolerance on the state learning and omitting the influence of the tail eigenvalues of the target states based on QPCA. Therefore,  
the efficient training condition of QSSM still applies to the low-rank state approximation learning by fixing a maximum scattering layer width.

\section{Quantum sequential scattering model}\label{sec:quantum_sequential_scattering_model}
The fundamental idea of state learning using the quantum sequential scattering model (QSSM) is to composite the target states by gradually aligning reduced density matrices of subsystems. The model diffuses the local quantum information into the global system, which can be considered a quantum analogy of the classical diffusion model. In contrast, the conventional QNN model handles the entire system at a time. We now present the overview of our QSSM with an efficient state learning algorithm.

Suppose we have access to the copies of an $n$-qubit pure target state $\rho = \ketbra{\phi}{\phi}$ from some other quantum instances. The target state can be represented in a system containing $n$ ordered quantum registers. Recalling $\rho_k$ as the reduced density matrix on the first $k$ registers, i.e., $\rho_k = \tr_{q_{k+1}:q_n}[\rho]$, our model aims to construct a purification  $\ket{\psi_k(\bm{\theta}_k)} = U_k(\bm{\theta}_k)\ket{\psi_{k-1}}$ of $\rho_k$ at the $k$-th learning step ($1\leq k\leq n$) by training the $k$-th scattering layer realized as a parameterised circuit $U_k(\bm{\theta}_k)$. Notice that the learning results from the previous step are naturally involved in the state $\ket{\psi_{k-1}}$ having all first $k$ registers aligned. 

The training of each layer is based on minimizing some adaptive cost functions, which in this work, we use the modified distance function of form~\ref{Eq: cost_func} where the $k$-th layer output state $\sigma_k(\bm{\theta}_k) = \tr_{q_{k+1}:q_n}[\ketbra{\psi_{k}(\bm{\theta}_k)}{\psi_{k}(\bm{\theta}_k)}]$. By hierarchically training the scattering layers until all registers are aligned, we could then construct the entire target through our trained quantum sequential scattering model.
\begin{algorithm}[H]
\caption{Quantum sequential scattering model for (pure) state learning}
\label{alg:algorithm}
\textbf{Require:}
Copies of the $n$-qubit target state $\rho = \ketbra{\phi}{\phi}$, Cost tolerance $\delta$.\\
\textbf{Ensure:}
The entire model has $n$ quantum registers as $q_1, q_2, \cdots, q_n$, and are initialized to $\ket{0}^{\otimes n}$.\\
\textbf{Parameter}: All layer parameters are randomly initialized regarding Uniform distribution of $[0,2\pi)$. Set $k=1$ and maximum layer width $w_{\max}$.
\begin{algorithmic}[1] %[1] enables line numbers
\STATE Update scattering layer width $w_k = k+1$, $\ket{\psi_k} = \ket{0}^{\otimes n}$.
\WHILE{$k\leq n$}
\IF {$k \leq \lfloor n/2\rfloor$}
\STATE $w_k = \min\{k+1, w_{\max}\}$.
\ELSIF{$k > \lfloor n/2\rfloor$}
\STATE $w_k = \min\{n-k+1, w_{\max}\}$.
\ENDIF
\STATE Apply $U_k(\bm{\theta}_k)$ to the quantum registers indexing $q_k$ to $q_{k+w_k -1}$, i.e., $q_k:q_{k+w_k -1}$.
\STATE Minimize $C_k(\bm{\theta}_k)$ via running classical training algorithm based on the analytic cost function and gradient $\nabla_{\bm{\theta}_k}C_k$ evaluations. The minimization stops until the cost difference reaches $\delta$.
\STATE $k = k+1$.
\STATE Update $\ket{\psi_k} = U_k(\bm{\theta}_k)\ket{\psi_{k-1}}$.
\ENDWHILE
\STATE Store all optimized $\bm{\theta}_1, \cdots,\bm{\theta}_n$ in classical memory.
\STATE \textbf{return} model reconstructed representation $\ket{\psi_n} = U_n\cdots U_1 \ket{0}^{\otimes n}\approx \ket{\phi}$.
\end{algorithmic}
\textbf{Output}: The trained QSSM as an approximate state generator $\mathbf{U} = U_n\cdots U_1$ of target $\ket{\phi}$.
\end{algorithm}
We summarize our quantum state learning algorithm via QSSM in Algorithm~\ref{alg:algorithm}. 
% A flowchart of QSSM state learning is illustrated in Fig.~\ref{Fig: QSSM_flowchart}. 

\begin{figure*}[t]
    \centering
    \includegraphics[width=1.0\linewidth]{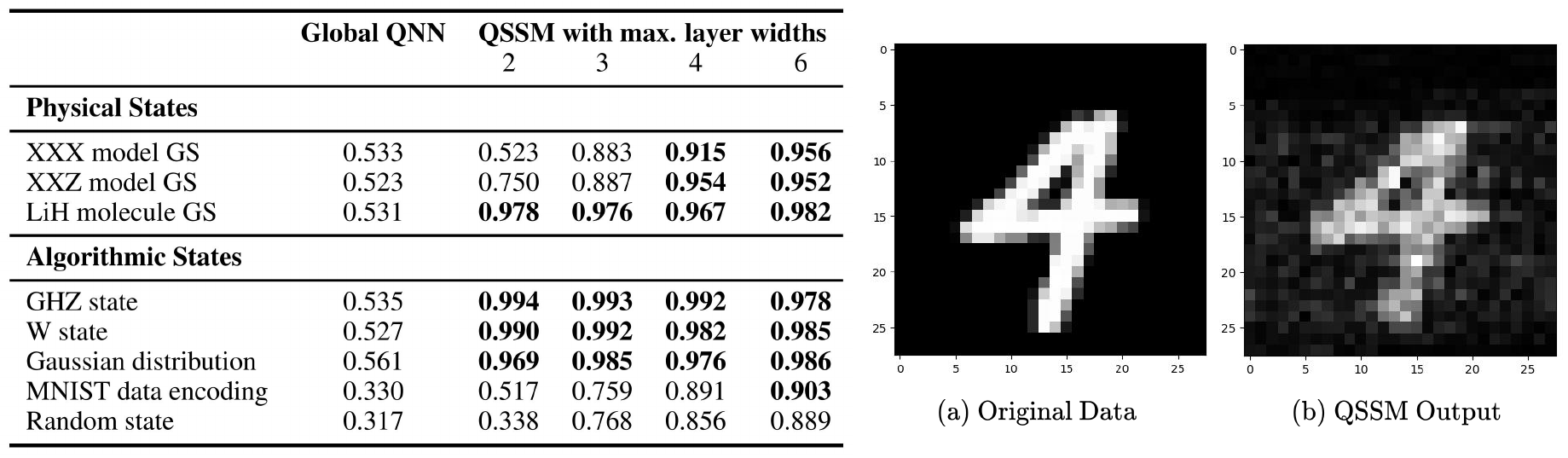}
    \caption{Effectiveness validation of QSSM in learning diverse $12$-qubit quantum states regarding their final state fidelities. On the right, we show the QSSM learnt state (b) from the MINST dataset concerning the original data (a) using amplitude encoding. With different maximum layer widths, our QSSM outperforms global QNN on state learning tasks.}
    \label{tab:truncation}
\end{figure*}

\subsection{Cost Function Evaluation}
As a hybrid quantum-classical model, we declare some details of the realization of the model in the following. For the adaptive $k$-th step cost function defined in~\eqref{Eq: cost_func}. By rearranging equation~\eqref{Eq: cost_func} as,
\begin{equation}
    C_k(\bm{\theta}_k) = \tr[\sigma^2_k(\bm{\theta}_k)] + \tr[\rho^2_k] - 2\tr[\sigma_k(\bm{\theta}_k)\rho_k],
\end{equation}
which is \textit{convex} according to Theorem 2.10 of~\cite{Carlen2009TRACEIA}.We chose this cost form since it can be efficiently evaluated on quantum hardware. The high-order state overlap terms involving $\tr[\rho^2]$ and $\tr[\rho\sigma]$ can be evaluated via \textit{swap test}~\cite{barenco1997stabilization}, which have been experimentally demonstrated on real quantum devices~\cite{Islam2015,Linke2018}. The training of the $k$-th layer can be described as finding the $k$-th step optimal parameters $\bm{\theta}_k^{opt}$ so that $C_k(\bm{\theta}_k^{opt})$ is minimized to approximately zero. To implement that, classical gradient-based and gradient-free methods, such as ADAM and COBYLA~\cite{kingma2014adam,powell1994direct}, can either be used during optimizations. 
Other metrics can also be employed in training procedures, and we left this aspect open for future research.

\subsection{Analytic Gradient Evaluation}
\label{subsec: cost and gradient}
Further, the analytical gradients of the cost function in~\eqref{Eq: cost_func} can be computed efficiently, making the gradient-based scheme a prospective candidate for the training processes. According to~\cite{Schuld2018,Mitarai2018,Ostaszewski2019,Wang2020a}. Suppose the $k$-th layer $U_k$ consists of the gates satisfying the \textit{parameter-shift rule}~\cite{Mitarai2018,Schuld2018} and contains $m$ trainable parameters. Each optimization iteration is driven by the estimations of cost gradient given by,
\begin{equation}
    \nabla_{\bm{\theta}_k} C_k({\bm{\theta}_k}) = \Big(\partial_1 C_k({\bm{\theta}_k}), \cdots, \partial_m C_k({\bm{\theta}_k})\Big),
\end{equation}
where $\partial_\mu := \frac{\partial}{\partial \theta_k^\mu}$ indicating the partial derivative with respect to a fixed $\theta^\mu_k$ in the $k$-th layer. In particular, we derive the analytic gradient of $C_k$ as follows,
\begin{equation}\label{Eq: exact form of partial mu cost}
    \partial_\mu C_k^* = \left\langle G_k^*\right\rangle_{(\theta_k^\mu)^* + \frac{\pi}{2}} - \left\langle G_k^*\right\rangle_{(\theta_k^\mu)^* - \frac{\pi}{2}}
\end{equation}
The symbol $*$ indicating the corresponding quantity evaluated at $\bm{\theta}_k = \bm{\theta}_k^*$. $G_k$ is a Hermitian operator involves both $\sigma_k$ and $\rho_k$ having an expression,
\begin{equation}
    G_k(\bm{\theta}_k) := \Delta_k(\bm{\theta}_k) \otimes \Gamma_k
\end{equation}
where $\Delta_k(\bm{\theta}_k) = \sigma_k(\bm{\theta}_k) - \rho_k$ representing the $k$-th step state difference between two density matrices; $\Gamma_k$ is the maximally mixed state $I/d$ where $I$ is the identity operator of dimension $d = 2^{w_k - 1}$. $\Gamma_k=1$ when $w_k = 1$. The bra-ket operation in the analytic form, $\langle A\rangle_{\alpha} = \braket{\psi_{k-1}}{U^\dag_k(\bm{\theta}_k)AU_k(\bm{\theta}_k)|\psi_{k-1}}$ for some Hermitian operator $A$ is evaluated at $\theta_k^\mu = \alpha$. This quantity of $G_k$ in~\eqref{Eq: exact form of partial mu cost} indicates the expectation value of $G_k$ regarding the $k$-th step variational ansatz $\ket{\psi_k}$ evaluated at $(\theta_k^\mu)^* \pm \pi/2$ where all other scattering layers remain unchanged. The detailed derivation of these definitions and forms can be found in Appendix~\ref{appendix:analytic_cost}.

Each partial derivative of $C_k$ at $\bm{\theta}_k^*$ can be explicitly determined by~\eqref{Eq: exact form of partial mu cost}, which can be efficiently computable via shifting the corresponding parameter and applying variational quantum eigensolver~\cite{Peruzzo2014}. The gradient-based optimization could be applied to the cost by specifically updating the parameters $\bm{\theta}_k$ in the $k$-th layer as,
\begin{equation}\label{Eq: updating step}
    \bm{\theta}_k \leftarrow \bm{\theta}_k^* - \eta \nabla_{\bm{\theta}_k} C_k(\bm{\theta}_k^*)
\end{equation}
where $\eta$ is the learning rate settled for the classical optimizers, defining the iteration step size. The cost function would converge to the optimal minimum by iterating the training processes. We then repeat the above procedures for each $k$-th layer to complete the model training with a final output circuit representation $\mathbf{U}(\bm{\theta}^{opt}) = U_n(\bm{\theta}_n^{opt})\cdots U_1(\bm{\theta}_1^{opt})$ to finish the state learning.

\section{Numerical Experiments}\label{sec:numerical_exp}
As described above, the adaptation of our quantum sequential scattering model indicates the underlying enhancement of information diffusion in quantum state learning. We now present numerical experiments to illustrate the effectiveness and trainability of QSSM.

% \begin{table*}[t]
% \centering
% \begin{tabular}{@{}lcccccc@{}}
% \toprule[1.5pt]
% & \multicolumn{1}{c}{} & \multicolumn{1}{c}{\textbf{Global QNN}} & \multicolumn{4}{c}{\textbf{QSSM with max. layer widths}}\\
% \multicolumn{3}{c}{} & 2 & 3 & 4 & 6\\ 
% \midrule[1.0pt]
% \multicolumn{7}{l}{\textbf{Physical States}}\\ 
% \midrule[1.0pt]
% \multicolumn{2}{l}{XXX model GS} & 0.533 & 0.523 & 0.883 & \textbf{0.915} & \textbf{0.956}\\
% \multicolumn{2}{l}{XXZ model GS} & 0.523 & 0.750 & 0.887 & \textbf{0.954} & \textbf{0.952}\\
% \multicolumn{2}{l}{LiH molecule GS} & 0.531 & \textbf{0.978} & \textbf{0.976} & \textbf{0.967} & \textbf{0.982}\\ 
% \midrule[1.0pt]
% \multicolumn{7}{l}{\textbf{Algorithmic States}}\\ 
% \midrule[1.0pt]
% \multicolumn{2}{l}{GHZ state} & 0.535 & \textbf{0.994} & \textbf{0.993} & \textbf{0.992} & \textbf{0.978}\\
% \multicolumn{2}{l}{W state} & 0.527 & \textbf{0.990}& \textbf{0.992}& \textbf{0.982}& \textbf{0.985}\\
% \multicolumn{2}{l}{Gaussian distribution} & 0.561 & \textbf{0.969} & \textbf{0.985} & \textbf{0.976} & \textbf{0.986}\\
% \multicolumn{2}{l}{MNIST data encoding} & 0.330 & 0.517 & 0.759 & 0.891 & \textbf{0.903}\\
% \multicolumn{2}{l}{Random state} & 0.317 & 0.338 & 0.768 & 0.856 & 0.889\\
% \bottomrule[1.5pt]
% \end{tabular}
% \caption{Flexibility of truncated-QSSM (t-QSSM) (noise-free) numerical simulations on learning both physical and algorithmic quantum states with increasing maximum layer widths $w_{\max}$.}
% \label{tab:truncation}
% \end{table*

\begin{figure*}[t]
    \centering
    \includegraphics[width=0.95\linewidth]{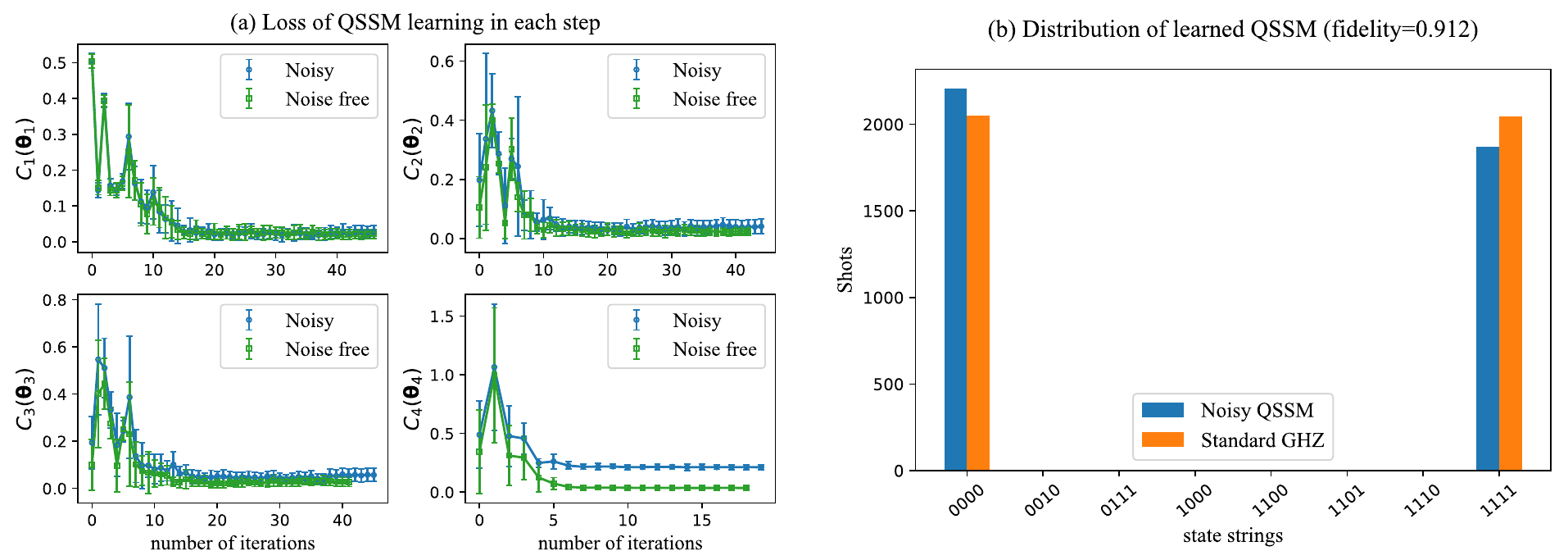}
    \caption{Noisy quantum simulation of QSSM for learning a $4$-qubit GHZ state. (a) Comparison of the variation of cost function noisy quantum simulation and noise-free simulation. For both cases, the optimization was processed via COBYLA optimizer~\cite{Gomez1994} on swap-test estimated cost values. (b) The distribution of measurement outcomes generated noise-freely from the state obtained by the noisy trained QSSM. The figure validates the efficacy and efficiency of QSSM  in noisy environments, consequently reinforcing our method's practical applicability.}
    \label{fig: noisy_sim}
\end{figure*}

We first conduct numerical simulations on QSSM for learning 12-qubit quantum states with physical or algorithmic meaning and compare our results with the performances from the conventional QNN model. The ground states from Heisenberg (XXX \& XXZ) models~\cite{takahashi1971one} and the LiH molecular model are pre-determined via the OpenFermion library developed by~\cite{mcclean2020openfermion}. For the Gaussian distribution and MNIST data learning experiments, the distribution and image data are normalized and mapped to the unit quantum state vectors of dimension $2^n$ via amplitude encoding~\cite{schuld2021supervised} with automatic padding of $0$'s filling out the extra grayscale pixels.  

In our numerical simulations involving the global QNN and the QSSM, we employ a general hardware efficient ansatz (HEA)~\cite{kandala2017hardware} of depth $d = 20$ with random initialized parameters for both the global model and each scattering layer in QSSM.  The optimization uses the ADAM optimizer with a learning rate of $0.1$ and cost tolerance $0.001$, spanning $200$ iterations. 

As shown in Fig.~\ref{tab:truncation}, comparing the outcomes with those of the global QNN, we discern clear advantages exhibited by QSSM, which consistently attains notably high fidelity in learning diverse quantum states. Conversely, the conventional model does not perform well, primarily due to the significantly decreased convergence speed during the training processes with a large number of qubits.

Besides, states with exponential growth in Schmidt ranks are not necessarily hard to learn. Only highly entangled states, e.g., random states and maximally entangled states (MES)~\cite{gisin1998bell}, are challenging for QSSM. Those with concentrated Schmidt coefficients, though owning large ranks, can be learnt up to a high fidelity~\cite{Liu2022a} with limited resources. 

In Table~\ref{tab:truncation}, we reasonably constrain the maximum scattering layer widths to some fixed values, which counterintuitively yield superior performance with smaller layer width. Larger values of $w_{\max}$, contrarily, decrease the QSSM performances of state learning. 
A plausible explanation for this phenomenon could be the over-parameterization and the mild BP effect during the training of the halved-dimensional scattering layers. Notably, learning random state undoubtedly obtains the worst learning results. 

We also examine the noise robustness of using QSSM to learn a $4$-qubit GHZ state on the IBMQ Qiskit simulator~\cite{Qiskit}. We build our noise model from single qubit and multi-qubit depolarizing channels (DCs) and thermal relaxation channels (TRCs)~\cite{georgopoulos2021modeling}. The error rate of DCs are set to $10^{-3}$, and the $T_1$, $T_2$ and gate time of TRCs are set to $1000 \ \mu\text{s}$, $100 \ \mu\text{s}$ and $1 \ \text{ns}$ respectively. 

At each step, we run the optimization of the QSSM circuit $20$ times in parallel and use the parameters that correspond to the lowest cost to update the circuit before going to the next step. This trick can significantly alleviate the randomness arising from sampling of bit strings in the measurement of quantum circuits. Shown in Fig.~\ref{fig: noisy_sim}, each learning step has cost converged well compared with the ideal training in (\ref{fig: noisy_sim}a). The final fidelity between the quantum state generated from QSSM and the true GHZ state could reach 91\%, giving almost the same statistical behaviours plotted from the sampling experiments (\ref{fig: noisy_sim}b). 

From the analytical description and numerical demonstration, we see that QSSM has the ability to learn arbitrary quantum states with high fidelity compared to the conventional model. The diffusion strategy only requires narrow circuits in learning quantum states that are weakly entangled, thus being extremely efficient in learning such a class of quantum states.

We then present the result to demonstrate Proposition~\ref{prop:main result on gradient of QSSM} by comparing the gradient variances of cost~\eqref{Eq: cost_func} as a function of the number of registers for QSSM and global QNN model. We typically investigate the values in the first step, the middle step ($\frac{n}{2}$-th step), and the last step of the QSSM learning procedure by looking into a single parameter $R_Z$ gate in the middle of each scattering layer. By assuming the two parts $U^{(k)}_\pm$ split by the $R_Z$ gate are deep enough to form local unitary 4-designs, we sample local Haar random unitaries~\cite{dankert2009exact} to simulate the behaviours of random initialization on $U^{(k)}_\pm$ and compute the gradient variances with respect to the parameter in $R_Z$. Similar experiments are performed for the conventional QNN model by sampling global Haar unitaries with a $R_Z$ gate sandwiched in. We target the GHZ state and the ground state of the Heisenberg model, as before, with maximum width $w_{\max}$ being 2 and 4, respectively. The variance values are computed from sampling 500 Haar unitary pairs for both cases.

\begin{figure*}[t]
    \centering
    \includegraphics[width = 0.9\textwidth]{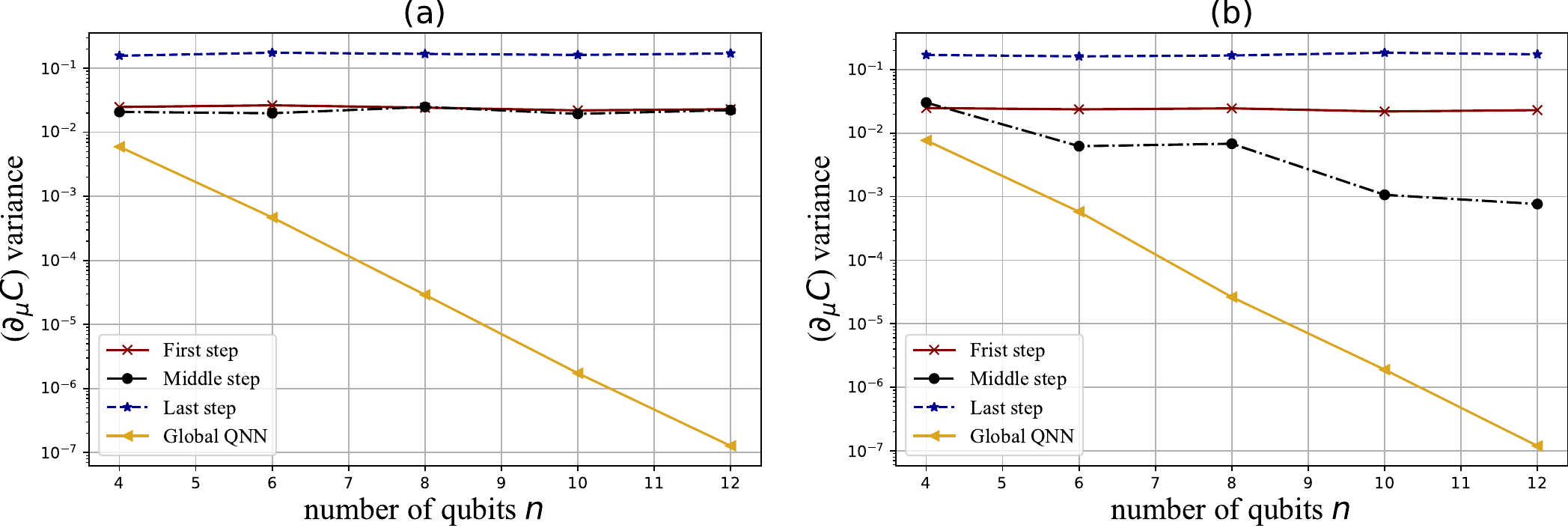}
    \caption{Comparison of the gradient variances as a function of the number of qubits on a semi-log plot from different steps in QSSM and global QNN computed by sampling Haar random unitaries. Panel (a) and (b) correspond to the learning of the GHZ state and the ground state of the Heisenberg model, respectively. The red, black and blue lines represent the gradient magnitudes of the first step, $\frac{n}{2}$-th step and the last step training, respectively, comparing with the global QNN results in yellow. Our method apparently outperforms conventional global QNN in terms of gradient variance scaling, indicating the absence of barren plateaus.}
    \label{fig: gradient analysis of QSSM}
\end{figure*}

As we can observe in Fig.~\ref{fig: gradient analysis of QSSM}. The variance of the gradient vanishes exponentially with the number of qubits when using the randomly initialized global QNNs. In contrast, QSSM demonstrates a constant scaling of variance magnitude. We note that there is a decay of the gradient variance of the middle step in panel (b). Nevertheless, this decay is caused by a constant factor $g(\rho_k)$ that originates from the nature of the physical system and does not exponentially influence the training processes.

\section{Conclusion and Discussion}
\label{sec: conclusion and future works}
In this paper, we have presented the development and application of the Quantum Sequential Scattering Model (QSSM) for quantum state learning. Our model is inspired by the classical diffusion model, which the designing of it involves quantum information theory and adaptive quantum machine learning techniques. Our theoretical analysis and numerical experiments demonstrate the superiority of the QSSM over conventional QNN approaches in terms of training speed and learning accuracy. In particular, the QSSM addresses the barren plateaus issues and provides an efficient solution to learning high-dimensional unknown quantum states based on sequentially learning the reduced target states. 

Moreover, We have analyzed the impact of increasing entanglement, a key property of quantum states, on the performance and efficiency of the QSSM. Our results show that the model can effectively handle polynomially increased entanglement, enabling us to learn complex quantum states accurately. Numerical demonstrations have shown out-performances for learning physical and algorithmic quantum states in terms of their rank-restricted approximations, indicating the broad applicability of QSSM state learning and the deep connection between state learning and quantum entanglement.

There are remaining issues of QSSM for future discussion. Different choices of scattering layers would influence the learning performance, which has to be exemplified. How to further improve the state fidelity provided the high fidelity state from QSSM could become a significant open question. Understanding and resolving the effect of over-parameterization from QSSM should be explained. A theoretical performance guarantee and the connection between scattering layer dilation and QSSM state learning information flow should be established for a complete story of truncated state learning. 
We also expect some extended applications of QSSM as a new quantum generative model instead of only state learning on near-term quantum devices.

\textbf{\textit{Acknowledgements.}}---M. J. and G. L. contributed equally to this work. Part of this work was done when M. J.,  G. L., and X. W. were at Baidu Research.

\bibliography{arxiv/arxiv}

%%%%%%%%%%%%%%%%%%%%%%%%%%%%%%%%%%%%%%%%%%%%%%%%%%%%%%%%%%%%%%%%%%%%%%%%%
% Supplementary Material

\clearpage
\appendix
\setcounter{subsection}{0}
\setcounter{table}{0}
\setcounter{figure}{0}
\numberwithin{equation}{section}
\renewcommand{\theproposition}{S\arabic{proposition}}
\renewcommand{\thelemma}{S\arabic{lemma}}
\renewcommand{\thedefinition}{S\arabic{definition}}
\renewcommand{\thefigure}{S\arabic{figure}}
\setcounter{equation}{0}
\setcounter{table}{0}
\setcounter{section}{0}
\newcounter{lemma}
\setcounter{lemma}{0}
\setcounter{proposition}{0}
\setcounter{definition}{0}
\setcounter{figure}{0}

\vspace{2cm}
\onecolumngrid
\vspace{2cm}

\begin{center}
\textbf{
{\Large{Supplementary Material for \\Quantum sequential scattering model for quantum state learning}}
}
\end{center}

% \vspace{2cm}
% \onecolumngrid
% \vspace{2cm}

\section{Preliminaries in  Quantum information}\label{appendix:quantum_info_basics}
In this appendix, we write more details on quantum computation and quantum information.

\subsection{Quantum computation and quantum information basics}
We use $\|\cdot\|_p$ to denote the $l_p$-norm for vectors and the Schatten-$p$ norm for matrices. The common-used linear algebra notations include complex conjugate transpose $A^\dagger$, the trace of matrix $\tr[A]$. The $\mu$-th component of the vector $\bm{\theta}$ is denoted as $\theta_\mu$. The derivative with respect to $\theta_\mu$ is then represented as $\partial_\mu := \frac{\partial}{\partial\theta_\mu}$. The big-O notation $\cO$ implies the asymptotic notation of upper bounds.

Quantum information is encoded and processed via the fundamental cells, namely, qubits, and described as quantum states. An $n$-qubit state can be mathematically represented by a $2^n \times 2^n$ positive semi-definite density matrix $\rho$, i.e., $\rho\succeq 0$ over the complex field and $\tr[\rho] = 1$. A pure state, in this formulation, satisfy $\rank{(\rho)} = 1$ and can be expressed in  \textit{Dirac bra-ket} notation as $\rho = \ketbra{\psi}{\psi}$ where $\ket{\psi}\in \CC^{2^n}$ denotes a \textit{Hilbert space} unit column vector with the corresponding \textit{dual vector} $\bra{\psi}^\dagger = \ket{\psi}$ and $\dagger$ denoting the complex conjugate transpose operation. A mixed state satisfies $\rank{(\rho)}>1$, and based on \textit{Spectral theorem}, it has a decomposition form $\rho = \sum_j p_j \ketbra{\psi_j}{\psi_j}$ where $p_j > 0$ denotes the probability of observing $\ketbra{\psi_j}{\psi_j}$ in $\rho$ and $\sum_j p_j = 1$. 

Based on \textit{Uhlmann's theorem}~\cite{nielsen2010quantum} for every mixed state $\rho$ acting as a linear operator on a Hilbert space $A$, there exists a purified state $\ket{AR}$ (i.e, pure state) in the composite system $AR$ such that $\tr_R[\ketbra{AR}{AR}] = \rho$, where $\tr_R[\cdot]$ denotes the partial trace operation tracing out the ancillary system $R$. The purification $\ket{AR}$ has a \textit{Schmidt decomposition} form $\ket{AR} = \sum_j \sqrt{p_j} \ket{\psi_j} \otimes \ket{j_R}$ for some orthonormal set $\ket{j_R}$ in $R$.

The partial trace operation in the above statement plays an important role in quantum computation and information. Given a composite quantum system described by a tensor product of Hilbert spaces, \(\mathcal{H}_A \otimes \mathcal{H}_B\), or simply denoted as $AB$, where \(\mathcal{H}_A\) and \(\mathcal{H}_B\) represent the Hilbert spaces of subsystems $A$ and $B$, respectively, the partial trace operation allows us to focus on subsystem $A$ while tracing out the degrees of freedom associated with subsystem $B$. The partial trace of an operator \(\rho\) with respect to subsystem B is denoted as \(\text{Tr}_B[\rho]\) and is defined as follows:
\[
\text{Tr}_B[\rho] = \sum_{i} (I_A \otimes \langle i|_B) \cdot \rho \cdot (I_A \otimes |i\rangle_B)
\]
Where \(I_A\) is the identity operator on \(\mathcal{H}_A\); \(|i\rangle_B\) forms an orthonormal basis for \(\mathcal{H}_B\) and \(\langle i|_B\) represents the conjugate transpose of \(|i\rangle_B\).

The evolution of a quantum state $\rho$ is realized by applying a series of quantum gates, which are mathematically described as unitary operators. The state $\rho'$ that undergoes transformation via a quantum gate $U$ can be obtained through direct matrix multiplication, expressed as $\rho' = U\rho U^\dagger$.
Common single-qubit gates include the Pauli rotations $\{R_P(\theta) = e^{-i\frac{\theta}{2}P} |P \in \{X,Y,Z\}\}$, which are in the matrix exponential form of Pauli matrices
\begin{equation*}
    X := 
    \begin{pmatrix}
        0 & 1\\
        1 & 0
    \end{pmatrix},
    Y := 
    \begin{pmatrix}
        0 & -i\\
        i & 0
    \end{pmatrix},
    Z :=
    \begin{pmatrix}
        1 & 0\\
        0 & -1
    \end{pmatrix}.
\end{equation*}

Common two-qubit gates include controlled-$X$ gate CX (or CNOT) $=I\oplus X$ and controlled-$Z$ gate CZ$=I\oplus Z$ where $\oplus$ denotes the direct sum operation. An $n$-qubit operator generally lives in the linear operator space $\cL(\CC^{2^n})$ over the complex field. Quantum measurements are then applied at the end of the quantum circuits, extracting classical information by projecting the quantum states onto its classical shadow.

\subsection{Fundamental of quantum neural networks}
In quantum machine learning, quantum neural networks (QNNs) are usually represented as parameterized unitaries consisting of a bunch of single-qubit rotation gates and several two-qubit gates, denoted as $\mathbf{U}(\bm{\theta})$ where $\bm{\theta}$ are the trainable parameters. The model is trained using a classical optimizer according to a minimization process on some cost function $C(\bm{\theta})$ based on the quantum measurement results. 

QNNs can be used to handle a variety of computational tasks, which is usually seen as a quantum version of classical neural networks. In the most general form, a QNN model can be expressed as $\mathbf{U}(\bm{\theta}) = \prod_{k=1}^M U_k(\bm{\theta}_k)$ for some sub-network layers $U_k(\bm{\theta}_k)$ where each layer can also be seen as a combination of parameterised circuits as $U_k(\bm{\theta}_k) = \prod_{j=1}^d U_j(\theta^{(k)}_j)W_j$, where $U_j(\theta^{(k)}_j) = e^{-ig_j\theta^{(k)}_j} $ is a parameterised gate with a Hermitian generator $g_j$. $W_j$ is usually non-parameterised, such as the networks of CNOT and CZ gates. The product $\prod_{k}$ here is, by default, in the increasing order from the right to the left in the above representations. 

The idea of quantum neural networks has obtained massive attention since its birth~\cite{toth1996quantum}. Various QNN architectures have been introduced to address a diverse range of computational challenges, spanning both classical and quantum problem domains~\cite{rebentrost2018quantum,zhao2019building,liu2013single,cong2019quantum,killoran2019continuous}, thereby pioneering an entirely novel realm of machine learning models. Recent literature focusing on the trainability theory of QNNs indicates a prospective direction for coping with barren plateaus by reducing the expressibility of QNN architectures~\cite{Cerezo2021, Liu2022}. Beyond that, some strategies have been proposed under certain conditions, for example, adopting clever initialization strategy~\cite{Grant2019, kulshrestha2022beinit}, using adaptive algorithms~\cite{Grimsley2019,Zhang2021,Skolik2021,Grimsley2022}, making parameterization generalization~\cite{volkoff2021large, friedrich2022avoiding} and choosing different cost forms and circuit architectures~\cite{Cerezo2021,Kieferova2021,Liu2022a}.

\section{Effectiveness of QSSM state learning}\label{appendix:effectiveness}
In this section, we give proof of the effectiveness of QSSM based on Schmidt decomposition, \textit{Uhlmann's theorem} and the properties of purification. 

\subsection{Degrees of freedom in Purification}
One of the implications of Uhlmann's theorem is that it ensures the degrees of freedom for quantum state purification~\cite{nielsen2010quantum}.
Purification is a commonly used mathematical procedure in quantum computing. For an arbitrary quantum state, its purification is not unique. However, we could bridge these purification states via unitary transformations, which we call freedom in purification.
\begin{lemma}\label{appendix lemma: freedom of purification}
    Let $\ket{\psi}$ and $\ket{\phi}$ be two purifications of a state $\rho$ acting on a composite system $AE$. Then there exists a unitary $U_{E}$ locally acting on $E$ s.t.,
    \begin{equation*}
        \ket{\psi} = (I_A\otimes U_{E})\ket{\phi}.
    \end{equation*}
\end{lemma}
The proof is simply inspired by the Schmidt decomposition. Let $\ket{\psi}$ and $\ket{\phi}$ be the purifications of $\rho$ acting on $AE$. Write the Schmidt decomposition of these two states,
\begin{equation*}
    \ket{\psi} = \sum_j \sqrt{\lambda_j} \ket{j_A}\ket{j_E} \quad \ket{\phi} = \sum_k \sqrt{\eta_k} \ket{k_A}\ket{k_E}.
\end{equation*}
Notice $\tr_E[\psi] = \rho = \tr_E[\phi]$, which then induces,
\begin{equation*}
    \sum_j\lambda_j \ket{j_A}\bra{j_A} = \sum_k \eta_k \ket{k_A}\bra{k_A}.
\end{equation*}
By linear algebra, we could easily extend both $\{\ket{j_A}\}_j$ and $\{\ket{k_E}\}_k$ to the basis set of $\cH_E$, via Gram-Schmidt method, and hence proves the existence of a unitary $U_E$ s.t,
\begin{equation*}
    U_E \ket{k_E} = \ket{j_E},
\end{equation*}
which is then substituted into the above equations to prove the lemma. Based on the freedom in purification, we could prove the lemma~\ref{lemma: effectiveness on restricted purification}, and therefore prove the effectiveness of our QSSM.
% Given a mixed state $\rho$ with a purification $\ket{AR}$, one can always find a local unitary $U_R$ acting on the ancillary system $R$ such that $\ket{AR'} = (I_A \otimes U_R)\ket{AR}$ forms another purification of $\rho$. Based on that, the theoretical guarantee of QSSM state learning is stated in Lemma~\ref{lemma: effectiveness on restricted purification}.
% \begin{lemma}\label{lemma: effectiveness on restricted purification}
%     Given a target state $\rho$ acting on system $A$ and $B$, and suppose its purification $\ket{\psi}$ on system $ABE$ where $E$ is an ancillary system, s.t.,
%     \begin{equation*}
%         \tr_{BE}[\ketbra{\psi}{\psi}] = \tr_{B}[\rho].
%     \end{equation*}
%     There always exists a local unitary $U_{BE}$ acting on the composite system $BE$, s.t.,
%     \begin{equation*}
%         \tr_{E}[(I_A\otimes U_{BE})\ketbra{\psi}{\psi}(I_A\otimes U_{BE}^\dag)] = \rho.
%     \end{equation*}
% \end{lemma}

\begin{lemma}\label{lemma: effectiveness on restricted purification}
    Given a target state $\rho$ acting on system $A$ and $B$, we suppose it can be purified on system $ABE$ where $E$ is an environment. For any pure state $\ket{\psi}$ acting on $ABE$, s.t.,
    \begin{equation*}
        \tr_{BE}[\ketbra{\psi}{\psi}] = \tr_{B}[\rho].
    \end{equation*}
    There always exists a local unitary $U_{BE}$, s.t.,
    \begin{equation*}
        \tr_{E}[(I_A\otimes U_{BE})\ketbra{\psi}{\psi}(I_A\otimes U_{BE}^\dag)] = \rho.
    \end{equation*}
\end{lemma}
From the definition, $\ketbra{\psi}{\psi}$ and $\rho$ have the same reduced state acting on $A$. Suppose the state $\ket{\phi}$ is the purification of $\rho$ on system $ABE$. Thus, it is also a purification of $\rho_A = \tr_B[\rho]$. We have $\ket{\phi}$ and $\ket{\psi}$ acting on the composite system $ABE$. By lemma \ref{appendix lemma: freedom of purification}, there exists a $U_{BE}$ s.t.,
\begin{equation*}
    \ketbra{\phi}{\phi} = (I_A\otimes U_{BE})\ketbra{\psi}{\psi}(I_A\otimes U_{BE}^\dag).
\end{equation*}
Now since $\ket{\phi}$ is the purification of $\rho$ we have,
\begin{equation*}
    \tr_{E}[(I_A\otimes U_{BE})\ketbra{\psi}{\psi}(I_A\otimes U_{BE}^\dag)] = \rho,
\end{equation*}
as required. Moreover, based on the Schmidt decomposition between $AB$ and $E$, the dimensionality of system $E$ clearly determines the maximum rank of the output states. For $\rank[\rho] = r$. It is sufficient and necessary to construct such a unitary $U_{BE}$ so that the last equation in lemma~\ref{lemma: effectiveness on restricted purification} can hold when $\dim[E] \geq \log_2 r$.

\subsection{Effectiveness proposition of QSSM}
\begin{figure*}[h]
    \centering
    \includegraphics[width=0.5\linewidth]{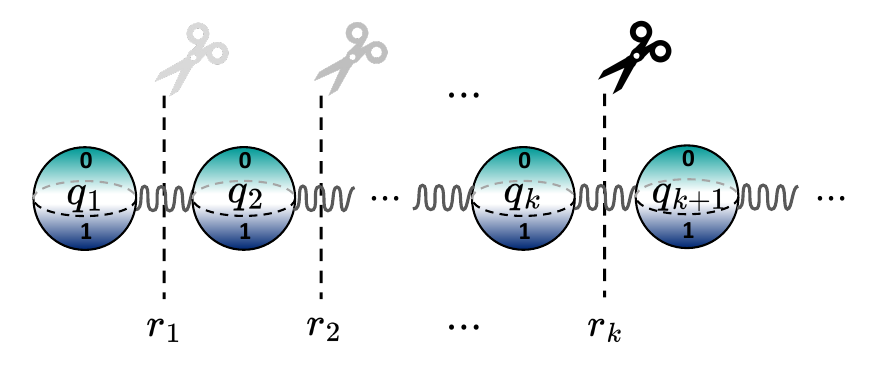}
    \caption{The sketch for illustrating the rank sequence of a given quantum state $\rho$.}
    \label{fig:rank_seq_def}
\end{figure*}

Before we move to the effectiveness proposition of QSSM state learning, we first define some symbols for a better layout of our demonstration of QSSM effectiveness. A $k$-th partition of $\rho$ separates the state into bipartite subsystems $\cA_k$ and $\Bar{\cA}_k$ covering the first $k$ qubits and the remaining, respectively, where $1\leq k\leq n$. For $k=n$, $\Bar{\cA}_k$ becomes trivial and $\cA_k = \CC^{2^n}$. We then could define the rank sequence of a given target state $\rho$ in the following sense. A sketch of this has been figured out in Fig.~\ref{fig:rank_seq_def}

\begin{definition}\label{def:rank sequence in app}
    Given an $n$-qubit quantum state $\rho$ represented by $n$ ordered quantum registers labeled as $q_1, q_2, \cdots, q_n$, denoting $\rho_k$ as the $k$-th reduced density matrix of the first $k$-register state, i.e., $\rho_k = \tr_{q_{k+1} : q_{n}}[\rho]$ for $1\leq k\leq n$ where the operation $\tr_{q_i:q_j}[\cdot]$ representing a partial tracing over registers $q_i$ to $q_j$, the (Schmidt) rank sequence of $\rho$ is an ordered list $\cR_\rho$,
    \begin{equation*}
        \cR_\rho = \{r_1, r_2, \cdots, r_{n-1}, r_{n}\},
    \end{equation*}
    where $r_k$ indicates  $\rank[\rho_k]$. In particular, if $\rho$ is pure, then $r_n = 1$ since $\rho$ can be represented as $\ketbra{\phi}{\phi}$ for some pure state vector $\ket{\phi}$.
\end{definition}
Here for clarification, by setting up the $k$-partition of $\rho$, $\cA_k$ contains the registers $q_1:q_k$ and $\Bar{\cA}_k$ contains the registers $q_{k+1}:q_n$ which is the reason why we use this notation to represent the corresponding partial trace operations. We are now ready to prove the effectiveness proposition of the main results.

\begin{proposition}\label{proposition: Truncation in app}[Effectiveness]
For a given $n$-qubit pure target state $\rho$ represented by $n$ ordered quantum registers $q_1, q_2, \cdots, q_n$, if the rank sequence of $\rho$ is $\cR_\rho = \{r_1, r_2, \cdots r_{n-1},r_n\}$. Then there exists a quantum algorithm~\ref{alg:algorithm}, based on QSSM, that could produce a state $\sigma$ exactly satisfying $\sigma = \rho$, if and only if the $k$-th scattering layer $U_k(\bm{\theta}_k)$ of QSSM has a width $w_k$ scales $\cO(\lceil \log_2 r_k\rceil)$.
\end{proposition}
To prove the above Proposition, we first suppose an $n$-qubit pure target $\rho = \ketbra{\phi}{\phi}$, and at the $k$-th step,
\begin{align*}\label{Eq: align_partial_state}
    \sigma_k = \tr_{\Bar{\cA}_k}[\ketbra{\psi_k}{\psi_k}]& = \tr_{\Bar{\cA}_k}[\rho] = \rho_k
\end{align*}
We call this the \textit{$k$-th perfect learning} condition of QSSM state learning. Then, by lemma~\ref{lemma: effectiveness on restricted purification}, there exists a local unitary such that,
\begin{align*}
    \tr_{\Bar{\cA}_{k+1}}[(I_k\otimes U_{k+1})\ketbra{\psi_k}{\psi_k} (I_k\otimes U_{k+1}^{\dagger})]& = \tr_{\Bar{\cA}_{k+1}}[\rho],
\end{align*}
where the existence of $U_{k+1}$ ensures the effectiveness of QSSM.
We call it a \textit{perfect learning} assumption of QSSM state learning if all the $k$-th perfect learning can be achieved.

Now, we are ready to deliver the proof of the effectiveness of QSSM. The proof assumes sufficient computational resources, ensuring perfect learning for each step's reduced target. 
% \renewcommand{\theproposition}{S\arabic{proposition}}
% \begin{proposition}
%     Given any target $n$-qubit pure state $\psi$, We claimed that our QSSM could ideally learn the state.
% \end{proposition}
We divide the entire learning task into three main stages based on the algorithm setup.

\textbf{(1)}, in the beginning, a state $\ket{0}$ is initialized for the model. We denote the step as $k = 1$ for learning the reduced state acting on $\cA_1$ of a single qubit. Notice that for any single-qubit state $\rho_1$ has an eigendecomposition,
\begin{equation*}
    \rho_1 = \lambda_1^{(1)} \ket{0^{(1)}}\bra{0^{(1)}} + \lambda_2^{(1)} \ket{1^{(1)}}\bra{1^{(1)}},
\end{equation*}
where the states $\ket{0^{(1)}}$ and $\ket{1^{(1)}}$ are not necessary the computational basis elements. There exists a purification unitary $U_{\cA_1\cA_2}$,
\begin{equation*}
    U_{\cA_1\cA_2}\ket{00} = \sqrt{\lambda_1^{(1)}} \ket{0^{(1)}}\ket{0^{(2)}} + \sqrt{\lambda_2^{(1)}} \ket{1^{(1)}}\ket{1^{(2)}}.
\end{equation*}
Such a unitary should have the following components. The rest of the matrix can be extended using the Gram-Schmidt process. We could write out the computational basis representation of $U_{\cA_1\cA_2}$,
\begin{equation*}
    [U_{\cA_1\cA_2}]_{mn} = 
    \begin{pmatrix}
        \sqrt{\lambda_1^{(1)}}\braket{00}{0^{(1)}0^{(2)}} + \sqrt{\lambda_2^{(1)}} \braket{00}{1^{(1)}1^{(2)}} & \cdots \\
        \sqrt{\lambda_1^{(1)}}\braket{01}{0^{(1)}0^{(2)}} + \sqrt{\lambda_2^{(1)}} \braket{01}{1^{(1)}1^{(2)}} & \cdots \\
        \sqrt{\lambda_1^{(1)}}\braket{10}{0^{(1)}0^{(2)}} + \sqrt{\lambda_2^{(1)}} \braket{10}{1^{(1)}1^{(2)}} & \cdots \\
        \sqrt{\lambda_1^{(1)}}\braket{11}{0^{(1)}0^{(2)}} + \sqrt{\lambda_2^{(1)}} \braket{11}{1^{(1)}1^{(2)}} & \cdots
    \end{pmatrix}.
\end{equation*}
\textbf{(2)}, now for $1 < k \leq \lceil n/2 \rceil$, by the assumption of ideal learning of state $\rho_{k-1}$, a purification, denoted as $\ket{\psi_{k-1}}$ of it would be imported from the $(k-1)$-th step. The reduced state $\rho_{k}$ would generally require at least $k$ extra ancillary qubits to be purified, which is why a width control $w_k = k + 1$ is settled in the worst case. Moreover, if the $\cR_{\rho}$ is given as above, the rank values give better choices of layer widths as $w_k = \min\{k+1, \lceil \log_2 r_k \rceil\}$.

Now suppose a purification $\ket{\phi_k}$ of $\rho_k$. Since $\Op{dim}(\ket{\psi_{k-1}}) \leq \Op{dim}(\ket{\phi_{k}})$, we could always extend $\ket{\psi_{k}}$ to $\ket{\Tilde{\psi_k}} = \ket{\psi_{k}}\ket{0}$ so that the result pure state lives in the same dimensional Hilbert as $\ket{\phi_{k}}$. We could observe $\ket{\Tilde{\psi_k}}$ and $\ket{\phi_k}$ are both purification of $\rho_{k-1}$. Based on the lemma \ref{lemma: effectiveness on restricted purification}, there exists $U_k$ acting on the qubits index from $k+1$ to $w_k+k$ s.t.,
\begin{equation*}
    \tr_{\Bar{\cA}_{k+1}}[(I_{\cA_{k-1}}\otimes U_k)\Tilde{\psi}_k(I_{\cA_{k-1}}\otimes U_k^\dag)] = \tr_{\Bar{\cA}_{k+1}}[\phi_k] = \rho_k.
\end{equation*}

\textbf{(3)}, at last, for $\lceil n/2 \rceil < k \leq n$. $\ket{\phi_k}$ becomes the pure state acting on the entire system of $n$ qubit registers. The imported purification $\ket{\psi_{k-1}}$ of $\rho_{k-1}$ is also a pure state of $n$ qubits. The result follows by applying the lemma \ref{lemma: effectiveness on restricted purification} again but with $w_k = \min\{n-k+1, \lceil\log_2 r_k\rceil\}$.

Above all, we have proven the effectiveness of QSSM. 
% The proof also encourages us to study situations when the sequential scattering unitary $U_k$ is not settled via the `doubling' strategy and enlighten the truncated version of QSSM.
One important point to note here is that the width of each scattering layer can be carefully settled concerning the rank of $\rho_k$ for $1\leq k< n$ in order to obtain the perfect learning.
% when the rank of $\rho_k$ for $1\leq k< n$ in the target is bounded above. An upper bound for each layer width $w_k$ can also lead to perfect learning. To better demonstrate the effectiveness, here we recall the definition of \textit{rank sequence}~\ref{def:rank sequence} of any quantum state characterizing the Schmidt rank distributions within the state. 
However, exactly constructing those purification unitaries using scattering layers $U_k(\bm{\theta}_k)$ is not possible. In reality, if each scattering layer of QSSM forms an approximate local unitary $t$-design for sufficient large positive integer $t$. Then, given enough time for training, the scattering layers would approximate these purification unitaries to arbitrarily high accuracy.

Further, the proposition identifies a group of quantum states that can be learned more efficiently using QSSM. One notable exemplar within this proposition is the $n$-qubit GHZ state.

\begin{remark}\label{Remark: GHZ}
An $n$-qubit GHZ state~\cite{greenberger1989going} has constant rank $r_k = 2$ for $1\leq k<n$. Hence, setting $w_k = 2 \ \forall k$ is sufficient to obtain perfect learning of QSSM state learning on GHZ state.
\end{remark}

The above phenomenon suggests a connection between the amount of entanglement within a target state and the sufficient widths $w_k$ to achieve perfect learning. The higher the ranks, the harder the target state could be learnt via QSSM. 

\section{Trainability and gradient analysis of QSSM}\label{appendix:trainability}
In this section, we give the proof for the proposition~\ref{prop:main result on gradient of QSSM} stated about the trainability of QSSM in this paper. We first recall some useful lemmas to make the proof easy to read and emphasize important intermediate results. The following lemmas were derived from the studies of unitary $t$-design. These were originally computed in~\cite{Cerezo2021}.
\begin{definition}\label{def: unitary t-design}
    A unitary $t$-design of dimension $d$~\cite{dankert2009exact} with respect to the Haar measure is defined as a finite set of unitaries $\{U_k\}_{k=1}^M$ on a $d$-dimensional Hilbert space such that,
    \begin{equation*}
        \frac{1}{M} \cdot \sum_{k=1}^M P_{(t,t)}(U_k) = \int_{\cU(d)} d\mu_{\text{Haar}}(U)P_{(t,t)}(U),
    \end{equation*}
    where $P_{(t,t)}(U)$ denotes a homogeneous polynomial of degree at most $t$ on the elements of $U$ and $U^\dag$.
\end{definition}
\begin{lemma}
\label{lemma: unitary design UAUBUCUD}
Suppose $X\subset \mathcal{U}(d)$ is unitary t-design, and $A,B,C,D$ are arbitrary linear operators. If $t\geq1$, then we have 
\begin{equation}
    \frac{1}{|X|}\sum_{U\in X}\tr[U^{\dagger}AUB] = \int_{\mathcal{U}(d)}\tr[U^{\dagger}AUB] d\eta(U) = \frac{\tr[A]\tr[B]}{d}
\end{equation}
If $t\geq2$, then we have
\begin{align}
    & \frac{1}{|X|}\sum_{U\in X}\tr[U^{\dagger}AUBU^{\dagger}CUD] = \int_{\mathcal{U}(d)}\tr[U^{\dagger}AUBU^{\dagger}CUD] d\eta(U) \label{equation-lemma 2 design formula}\\ 
    &= \frac{\tr[A]\tr[C]\tr[BD]+\tr[AC]\tr[B]\tr[D]}{d^2-1}-\frac{\tr[AC]\tr[BD] + \tr[A]\tr[B]\tr[C]\tr[D]}{d(d^2-1)}\label{equation-lemma 2 design value}
\end{align}
\end{lemma}

\begin{lemma}
\label{lemma: integral trace ab trace cd}
Suppose $A,B,C,D$ are arbitrary linear operators. Then,
\begin{align*}
    \int_{\mathcal{U}(d)}\tr[UAU^{\dag}B]\tr[UCU^{\dag}D]d\eta(U) &= \frac{1}{d^2-1}(\tr[A]\tr[B]\tr[C]\tr[D]+\tr[AC]\tr[BD])\\
    &-\frac{1}{d(d^2-1)}(\tr[AC]\tr[B]\tr[D]+\tr[A]\tr[C]\tr[BD])
\end{align*}
\end{lemma}

\begin{lemma}
\label{lemma: bipartite_I_tensor_U}
Let $\mathcal{H} = \mathcal{H_A}\otimes \mathcal{H_B}$ be a bipartite Hilbert space of dimension $d = d_Ad_B$, and for arbitrary linear operators $M,N: \mathcal{H}\rightarrow \mathcal{H}$, we have
\begin{align*}
    \int_{\mathcal{U} (d_B)} d\eta(U) (I_A\otimes U)M(I_A\otimes U^{\dag})N = \frac{\tr_B[M]\otimes I_B}{d_B}N,
\end{align*}
and
\begin{align*}
    \int_{\mathcal{U}(d_B)} d\eta(U) \tr[(I_A\otimes U)M(I_A\otimes U^{\dag})N ]= \frac{\tr[\tr_B[M]\tr_B[N]]}{d_B}.
\end{align*}
\end{lemma}

\begin{lemma}
\label{lemma: bipartite_I_tensor_U_discrete}
Let $\mathcal{H} = \mathcal{H_A}\otimes \mathcal{H_B}$ be a bipartite Hilbert space of dimension $d = d_Ad_B$ $(d = 2^n,d_A = 2^{n'})$, and for arbitrary linear operators $M,N,U: \mathcal{H}\rightarrow \mathcal{H}$, we have
\begin{align*}
    \tr[(I_A\otimes U)M(I_A\otimes U^{\dag})N] = \sum_{p,q}\tr[UM_{qp}U^{\dag}N_{pq}],
\end{align*}
where the summation runs over all bitstrings of length $n'$, and where
\begin{align*}
    M_{qp} &= \tr_A[(\ketbra{p}{q}\otimes I)M]\\
    N_{pq} &=\tr_A[(\ketbra{q}{p}\otimes I)N].
\end{align*}
\end{lemma}

With these lemmas, we can now start our proof by directly calculating the variance of gradients. The whole proof includes three parts indicating the gradient magnitude of different stages in the algorithm.

\subsection{Trainability of the last layer}
\begin{proposition}\label{proposition: last step gradient}
For a $n$-qubit target state $\rho$, assume we start from the $\hat{\sigma}$ such that $\tr_n[\rho] = \tr_n[\hat{\sigma}]$, where $\tr_n[\rho]$ denotes partial trace over the last qubit of the state. And if the circuit is only acting on the last qubit and forms a 2-design, then $\EE[\partial_\mu C_n] = 0$ and the variance $\var[\partial_\mu C_n]\in[\frac{16}{27},\frac{8}{9}]$.
\end{proposition}
The proof is given by the following, suppose the output state is $\sigma$, then the cost function is
\begin{equation*}
    C_n(\bm{\theta}) = \tr[(\rho-\sigma(\bm{\theta}))(\rho-\sigma(\bm{\theta}))^{\dagger}].
\end{equation*}
With a similar notation used in McClean's paper~\cite{McClean2018}, we can use $U$ to denote the unitary representation of circuits. And we can write it as $U = U_+e^{-i\theta_{\mu} H}U_-$, where $H$ denotes the hermitian operator and in most cases it will be the Pauli matrices, and they are traceless. Since $\tr_n[\rho] = \tr_n[\hat{\sigma}]$, we have
\begin{equation*}
    \hat{\sigma}  = (I_A\otimes V_B)\rho(I_A\otimes V_B^{\dagger}).
\end{equation*}
where $V$ is a fixed unitary and system $A$ denotes the first $n-1$ qubits and the system $B$ denotes the last qubit. So $d_A = 2^{n-1}$ and $d_B=2$. For simplicity, we will hide the subscript in the following proof.

We then arrive at
\begin{equation*}
    \sigma = (I \otimes U V)\rho(I\otimes V^\dag U^\dag).
\end{equation*}

Next, we compute the partial derivative of $C_n$ w.r.t the $k$-th parameter. Notice that the trace is linear, the derivative operation could pass through the trace and hence we obtain,
\begin{equation*}
    \partial_\mu C_n = \partial_\mu (\tr(\rho^2 + \sigma^2 - 2(\rho\sigma)) = -2\tr(\rho\partial_\mu(\sigma)),
\end{equation*}

Now We start by calculating the mean of gradients, expanding the expression for $\sigma$, we could find,
\begin{equation*}
    \partial_\mu C_n = -2\tr\left[\rho \left((I\otimes(\partial_\mu U) V)\rho(I\otimes V^{\dag} U^\dag) + (I\otimes UV)\rho(I\otimes V^{\dag}(\partial_\mu U^\dag))\right)\right],
\end{equation*}
by the chain rule of derivative. 
Since $U = U_+e^{-i\theta_{\mu} H}U_-$, we could compute the derivatives as,
\begin{equation*}
    \begin{cases}
    \partial_\mu U = -i U_+ e^{-i\theta_{\mu} H} H U_-\\
    \partial_\mu U^\dag = i U_-^\dag H e^{i\theta_{\mu} H} U_+^\dag.
    \end{cases}
\end{equation*}
For convenient, we define $\Tilde{U}_+=U_+ e^{-i\theta_{\mu} H}$. Substituting the above into the expression of cost derivative to achieve,
\begin{equation*}
    \partial_\mu C_n = 2i\tr\left[\rho \left((I\otimes \Tilde{U}_+ H U_- V)\rho(I\otimes V^{\dag}U^\dag) - (I\otimes UV)\rho(I\otimes V^{\dag}U_-^\dag H \Tilde{U}_+^\dag)\right)\right].
\end{equation*}
Now we expand $U = \Tilde{U}_+ U_-$, and assume the $\Tilde{U}_- = U_-V$
\begin{align*}
    \partial_\mu C_n &= 2i\tr\left[\rho \left((I\otimes \Tilde{U}_+ H U_- V)\rho(I\otimes V^{\dag}U_{-}^\dag\Tilde{U}_{+}^{\dag}) - (I\otimes \Tilde{U}_+ U_-V)\rho(I\otimes V^{\dag}U_-^\dag H \Tilde{U}_+^\dag)\right)\right]\\
    &= 2i\tr\left[\rho \left((I\otimes \Tilde{U}_+ H \Tilde{U}_-)\rho(I\otimes \Tilde{U}_-^{\dag}\Tilde{U}_{+}^{\dag}) - (I\otimes \Tilde{U}_+ \Tilde{U}_-)\rho(I\otimes \Tilde{U}_-^{\dag} H \Tilde{U}_+^\dag)\right)\right]\\
    &= 2i\tr\left[(I\otimes \Tilde{U}_{+}^{\dag})\rho(I\otimes \Tilde{U}_{+}) [I\otimes H, (I\otimes \Tilde{U}_{-})\rho(I\otimes \Tilde{U}_{-}^{\dag}) ]\right].
\end{align*}
where the $[A,B] = AB - BA$ denotes the commutator notation.
Denote the commutator $[I\otimes H, (I\otimes \Tilde{U}_{-})\rho(I\otimes \Tilde{U}_{-}^{\dag})]$ by $T_-$, thus we have
\begin{align*}
    \partial_\mu C_n &= 2i\tr\left[ (I\otimes \Tilde{U}_{+}^{\dag})\rho(I\otimes \Tilde{U}_{+})T_- \right].
\end{align*}
Then we integrate over $\Tilde{U}_+$ by using the lemma~\ref{lemma: bipartite_I_tensor_U},
\begin{align*}
    \mathbb{E}[\partial_\mu C_n] &= 2i \frac{\tr[\tr_B[\rho]\tr_B[T_-]]}{d_B}\\
    & = i\tr[\tr_B[\rho]\tr_B[T_-]].
\end{align*}
We can write the $\rho$ as
\begin{equation*}
    \rho = \sum_{i,j}\ketbra{i}{j}_A\otimes X_{i,j}.
\end{equation*}
thus lead to
\begin{align}
    \tr_B[T_-] &= \tr_B[[I\otimes H, (I\otimes \Tilde{U}_{-})\rho(I\otimes \Tilde{U}_{-}^{\dag})]] \nonumber\\
    &=\sum_{i,j}\tr_B[[I\otimes H, (I\otimes \Tilde{U}_{-})(\ketbra{i}{j}_A\otimes X_{i,j}) (I\otimes \Tilde{U}_{-}^{\dag})]]\nonumber\\
    &=\sum_{i,j}\tr_B[\ketbra{i}{j}\otimes H\Tilde{U}_{-}X_{i,j}\Tilde{U}_{-}^{\dag} - \ketbra{i}{j}\otimes \Tilde{U}_{-}X_{i,j}\Tilde{U}_{-}^{\dag}H]\nonumber\\
    &= \sum_{i,j}\ketbra{i}{j}(\tr[ H\Tilde{U}_{-}X_{i,j}\Tilde{U}_{-}^{\dag}]-\tr[\Tilde{U}_{-}X_{i,j}\Tilde{U}_{-}^{\dag}H]) \nonumber\\
    & =0 \label{equation: trace T_- = 0}.
\end{align}
Therefore, we have
\begin{align*}
    \mathbb{E}[\partial_\mu C_n] &= 0.
\end{align*}
The mean of gradients is $0$. Based on the fact that the mean of gradients is $0$, we then only need to consider the $\mathbb{E}[(\partial_\mu C_n)^2]$ in order to determine the variance.
\begin{align*}
    \var[\partial_\mu C_n] = \mathbb{E}[(\partial_\mu C_n)^2] = -4 \mathbb{E}_{\Tilde{U}_{+},\Tilde{U}_{-}}\left[(\tr[ (I\otimes \Tilde{U}_{+}^{\dag})\rho(I\otimes \Tilde{U}_{+})T_- ])^2\right].
\end{align*}
Using lemma~\ref{lemma: bipartite_I_tensor_U_discrete}, we have
\begin{align*}
    \mathbb{E}_{\Tilde{U}_{+},\Tilde{U}_{-}}\left[(\tr[ (I\otimes \Tilde{U}_{+}^{\dag})\rho(I\otimes \Tilde{U}_{+})T_- ])^2\right] &= \mathbb{E}_{\Tilde{U}_{+},\Tilde{U}_{-}}\left[(\sum_{p,q}\tr[\Tilde{U}_+\rho_{qp}\Tilde{U}_+^{\dag}T_{-pq}])(\sum_{m,n}\tr[\Tilde{U}_+\rho_{nm}\Tilde{U}_+^{\dag}T_{-mn}]) \right]\\
    & = \mathbb{E}_{\Tilde{U}_{+},\Tilde{U}_{-}}\left[\sum_{p,q,m,n}\tr[\Tilde{U}_+\rho_{qp}\Tilde{U}_+^{\dag}T_{-pq}]\tr[\Tilde{U}_+\rho_{nm}\Tilde{U}_+^{\dag}T_{-mn}] \right]\\
    &= \sum_{p,q,m,n}\mathbb{E}_{\Tilde{U}_{+},\Tilde{U}_{-}}\left[\tr[\Tilde{U}_+\rho_{qp}\Tilde{U}_+^{\dag}T_{-pq}]\tr[\Tilde{U}_+\rho_{nm}\Tilde{U}_+^{\dag}T_{-mn}] \right].
\end{align*}
Then, according to lemma~\ref{lemma: integral trace ab trace cd}
\begin{align}
    &\sum_{p,q,m,n}\mathbb{E}_{\Tilde{U}_{+},\Tilde{U}_{-}}\left[\tr[\Tilde{U}_+\rho_{qp}\Tilde{U}_+^{\dag}T_{-pq}]\tr[\Tilde{U}_+\rho_{nm}\Tilde{U}_+^{\dag}T_{-mn}] \right]\nonumber\\
    =& \sum_{p,q,m,n}\mathbb{E}_{\Tilde{U}_{-}}( 
    \frac{1}{d_B^2-1}(\tr[\rho_{qp}]\tr[T_{-pq}]\tr[\rho_{nm}]\tr[T_{-mn}]+\tr[\rho_{qp}\rho_{nm}]\tr[T_{-pq}T_{-mn}]) \nonumber\\
    &- \frac{1}{d_B(d_B^2-1)}(\tr[\rho_{qp}\rho_{nm}]\tr[T_{-pq}]\tr[T_{-mn}]+\tr[\rho_{qp}]\tr[\rho_{nm}]\tr[T_{-pq}T_{-mn}])). \label{equation: variance of gradients step 2}
\end{align}
Since
\begin{align}
    \tr[\rho_{qp}] &=  \tr[\tr_A[(\ketbra{p}{q}\otimes I)\rho]]\nonumber\\
    &=\tr[(\ketbra{p}{q}\otimes I)\rho]\nonumber\\
    &=\tr[\ketbra{p}{q}\tr_B[\rho]]\nonumber\\
    &=\bra{q}\tr_B[\rho]\ket{p}, \label{equation: trace rho_qp}
\end{align}
and
\begin{align}
    \tr[T_{-pq}] &=  \tr[\tr_A[(\ketbra{q}{p}\otimes I)T_-]]\nonumber \\
    &=\tr[(\ketbra{q}{p}\otimes I)T_{-}]\nonumber\\
    &=\tr[\ketbra{q}{p}\tr_B[T_-]] \nonumber\\
    &= 0\label{equation: trace T_pq}.
\end{align}
where the Eq.~\ref{equation: trace T_pq} holds because of Eq.~\ref{equation: trace T_- = 0}.
 
Thus the Eq.~\ref{equation: variance of gradients step 2} can be simplified as 
\begin{align*}
    &\sum_{p,q,m,n}\mathbb{E}_{\Tilde{U}_{+},\Tilde{U}_{-}}\left[\tr[\Tilde{U}_+\rho_{qp}\Tilde{U}_+^{\dag}T_{-pq}]\tr[\Tilde{U}_+\rho_{nm}\Tilde{U}_+^{\dag}T_{-mn}] \right] \nonumber\\
    =& \sum_{p,q,m,n}\mathbb{E}_{\Tilde{U}_{-}}( 
    \frac{1}{d_B^2-1}\tr[\rho_{qp}\rho_{nm}]\tr[T_{-pq}T_{-mn}] - \frac{1}{d_B(d_B^2-1)}\tr[\rho_{qp}]\tr[\rho_{nm}]\tr[T_{-pq}T_{-mn}])\\
    =&\sum_{p,q,m,n} \mathbb{E}_{\Tilde{U}_{-}} \left(\frac{1}{d_B(d_B^2-1)}\tr[T_{-pq}T_{-mn}](d_B\tr[\rho_{qp}\rho_{nm}] - \tr[\rho_{qp}]\tr[\rho_{nm}])\right)\\
    =&\sum_{p,q,m,n} \frac{1}{d_B(d_B^2-1)}(d_B\tr[\rho_{qp}\rho_{nm}] - \tr[\rho_{qp}]\tr[\rho_{nm}])\mathbb{E}_{\Tilde{U}_{-}} \left(\tr[T_{-pq}T_{-mn}]\right). \label{equation: variance of gradients step 3}
\end{align*}
We now need to evaluate the other integral w.r.t $\Tilde{U}_-$. A simplification can be first done by noticing,
\begin{align*}
    T_{-pq} &= \tr_A[(\ketbra{q}{p}\otimes I)T_-]\\
    &= \tr_A[I\otimes H, (I\otimes\Tilde{U}_-)(\ketbra{q}{p}\otimes I)\rho(I\otimes\Tilde{U}_-^\dag)]\\
    &= [H, \Tilde{U}_- \tr_A[\ketbra{q}{p}\otimes I)\rho]\Tilde{U}_-^\dag]\\
    &= [H, \Tilde{U}_- \rho_{pq}\Tilde{U}_-^\dag],
\end{align*}
since $\ketbra{p}{q}\otimes I$ commutes with other operators. Therefore,
\begin{align*}
    \tr[T_{-pq}T_{-mn}] &= \tr[[H, \Tilde{U}_- \rho_{pq}\Tilde{U}_-^\dag][H, \Tilde{U}_- \rho_{mn}\Tilde{U}_-^\dag]]\\
    &= 2\tr[H\Tilde{U}_-\rho_{pq}\Tilde{U}_-^\dag H \Tilde{U}_- \rho_{mn} \Tilde{U}_-^\dag] - \tr[\Tilde{U}_-\rho_{pq}\rho_{mn}\Tilde{U}_-^{\dagger}H^2]-\tr[\Tilde{U}_-\rho_{mn}\rho_{pq}\Tilde{U}_-^{\dagger}H^2].
\end{align*}
So according to lemma~\ref{lemma: unitary design UAUBUCUD},
\begin{align*}
    &\mathbb{E}_{\Tilde{U}_{-}} \left(\tr[T_{-pq}T_{-mn}]\right) \\
    =& \frac{2}{d_B^2-1}(\tr[\rho_{pq}]\tr[\rho_{mn}]\tr[H^2]+\tr[\rho_{pq}\rho_{mn}]\tr^2[H]) \nonumber\\
    &-\frac{2}{d_B(d_B^2-1)}(\tr[\rho_{pq}\rho_{mn}]\tr[H^2]
    + \tr[\rho_{pq}]\tr[\rho_{mn}]\tr^2[H])-\frac{2}{d_B}\tr[\rho_{pq}\rho_{mn}]\tr[H^2]\\
    =&\frac{-2}{d_B(d_B^2-1)}(d_B\tr[\rho_{pq}\rho_{mn}]-\tr[\rho_{pq}]\tr[\rho_{mn}])(d_B\tr[H^2]-\tr^2[H])\\
    =&\frac{-2}{(d_B^2-1)}\tr[H^2](d_B\tr[\rho_{pq}\rho_{mn}]-\tr[\rho_{pq}]\tr[\rho_{mn}]).
\end{align*}
Then, we go back to Eq.~\ref{equation: variance of gradients step 3},
\begin{align*}
     &\sum_{p,q,m,n}\mathbb{E}_{\Tilde{U}_{+},\Tilde{U}_{-}}\left[\tr[\Tilde{U}_+\rho_{qp}\Tilde{U}_+^{\dag}T_{-pq}]\tr[\Tilde{U}_+\rho_{nm}\Tilde{U}_+^{\dag}T_{-mn}] \right] \nonumber\\
     =&\sum_{p,q,m,n} \frac{-2}{d_B(d_B^2-1)^2}\tr[H^2](d_B\tr[\rho_{qp}\rho_{nm}] - \tr[\rho_{qp}]\tr[\rho_{nm}])(d_B\tr[\rho_{pq}\rho_{mn}]-\tr[\rho_{pq}]\tr[\rho_{mn}]).
\end{align*}

First, we look at the $\tr[\rho_{qp}\rho_{nm}]$
\begin{align*}
    \tr[\rho_{qp}\rho_{nm}]
    &= \tr[\tr_A[(\ketbra{p}{q}\otimes I)\rho]\tr_A[(\ketbra{m}{n}\otimes I)\rho]]\\
    &= \tr[\sum_i \left( \bra{i}\otimes I \left( (\ketbra{p}{q}\otimes I)\rho \right) \ket{i}\otimes I \right) \sum_j (\bra{j}\otimes I (\ketbra{p}{q}\otimes I)\rho \ket{j}\otimes I) ]\\
    &=\tr[(\bra{q}\otimes I) \rho (\ketbra{p}{n}\otimes I )\rho (\ket{m}\otimes I)]\\
    &=\tr[\bra{q}\tr_B[\rho (\ketbra{p}{n}\otimes I) \rho]\ket{m}]\\
    &=\bra{q}\tr_B[\rho (\ketbra{p}{n}\otimes I) \rho]\ket{m}.
\end{align*}

Then,
\begin{align*}
    &\sum_{p,q,m,n}\tr[\rho_{qp}\rho_{nm}]\tr[\rho_{pq}\rho_{mn}]\\
    =& \sum_{p,q,m,n}\bra{q}\tr_B[\rho (\ketbra{p}{n}\otimes I) \rho]\ket{m}\bra{m}\tr_B[\rho(\ketbra{n}{p}\otimes I)\rho]\ket{q}\\
    =&\sum_{p,n}\tr\left[\tr_B[\rho (\ketbra{p}{n}\otimes I)\rho]\tr_B[\rho(\ketbra{n}{p}\otimes I)\rho]\right].
\end{align*}

Suppose the Schmidt decomposition of $\ket{\phi}$ is
\begin{align}
    \ket{\phi} = \sum_{k}\lambda_k \ket{u_k}_A\ket{v_k}_B.
\end{align}
where $\{\ket{u_k}\}$ are orthogonal basis on the system A and  $\{\ket{v_k}\}$ are orthogonal basis on the system B. 
Therefore, we can write the $\rho$ as
\begin{align}
    \rho = \sum_{i,j}\lambda_i\lambda_j \ketbra{u_i}{u_j}\otimes \ketbra{v_i}{v_j}.
\end{align}
We can expand the $\rho$ in $\tr_B[\rho (\ketbra{p}{n}\otimes I)\rho]$
\begin{align*}
    &\tr_B[\rho (\ketbra{p}{n}\otimes I)\rho] \\
    =& \tr_B[(\sum_{i,j}\lambda_i\lambda_j\ketbra{u_i}{u_j}\otimes \ketbra{v_i}{v_j}) (\ketbra{p}{n}\otimes I) (\sum_{k,l}\lambda_k\lambda_l\ketbra{u_k}{u_l}\otimes \ketbra{v_k}{v_l})]\\
    =&\sum_{i,j,k,l}\lambda_i\lambda_j\lambda_k\lambda_l \tr_B[\ketbra{u_i}{u_j}\ketbra{p}{n}\ketbra{u_k}{u_l}\otimes \ketbra{v_i}{v_j}\ketbra{v_k}{v_l}]\\
    =& \sum_{i,j}\lambda_i^2\lambda_j^2 \ketbra{u_i}{u_j}\ketbra{p}{n}\ketbra{u_j}{u_i}.
\end{align*}
Thus, we arrive at
\begin{align*}
    &\sum_{p,n}\tr\left[\tr_B[\rho (\ketbra{p}{n}\otimes I)\rho]\tr_B[\rho(\ketbra{n}{p}\otimes I)\rho]\right]\\
    =&\sum_{p,n}\tr[(\sum_{i,j}\lambda_i^2\lambda_j^2 \ketbra{u_i}{u_j}\ketbra{p}{n}\ketbra{u_j}{u_i})(\sum_{k,l}\lambda_k^2\lambda_l^2 \ketbra{u_k}{u_l}\ketbra{p}{n}\ketbra{u_k}{u_l})]\\
    =&\sum_{p,n}\tr[\sum_{i,j,k,l}]\lambda_i^2\lambda_j^2\lambda_k^2\lambda_l^2 \ketbra{u_i}{u_j}\ketbra{p}{n}\ketbra{u_j}{u_i}\ketbra{u_k}{u_l}\ketbra{n}{p}\ketbra{u_l}{u_k}\\
    =&\sum_{p,n}\sum_{i,j,l}\lambda_i^4\lambda_j^2\lambda_l^2 \tr[\bra{u_j}\ketbra{p}{n}\ketbra{u_j}{u_l}\ketbra{n}{p}\ket{u_l}]\\
    =&\sum_{i,j,l}\lambda_i^4\lambda_j^2\lambda_l^2 \tr[\tr[\ketbra{u_l}{u_j}]\tr[\ketbra{u_j}{u_l}]]\\
    =&\sum_{i,j}\lambda_i^4\lambda_j^4\\
    =& (\sum_i \lambda_i^4)^2.
\end{align*}
Then we look at the $\tr[\rho_{qp}]\tr[\rho_{pq}]\tr[\rho_{mn}]\tr[\rho_{nm}]$,
\begin{align*}
    &\sum_{p,q,m,n}\tr[\rho_{qp}]\tr[\rho_{pq}]\tr[\rho_{mn}]\tr[\rho_{nm}]\\
    =&\sum_{p,q,m,n}\bra{q}\tr_B[\rho]\ket{p}\bra{p}\tr_B[\rho]\ket{q}\bra{m}\tr_B[\rho]\ket{n}\bra{n}\tr_B[\rho]\ket{m}\\
    =&\tr[\tr_B[\rho]\tr_B[\rho]]\tr[\tr_B[\rho]\tr_B[\rho]]\\
    =&(\tr[\tr_B[\rho]\tr_B[\rho]])^2\\
    =&(\sum_i \lambda_i^4)^2.
\end{align*}
Now, we look at the $\tr[\rho_{qp}\rho_{nm}]\tr[\rho_{pq}]\tr[\rho_{mn}]$
\begin{align}
    \tr[\rho_{qp}\rho_{nm}] &= \bra{n}\tr_B[\rho (\ketbra{m}{q}\otimes I)\rho] \ket{p}\\
    &= \sum_{i,j}\lambda_i^2\lambda_j^2 \bra{n}\ketbra{u_i}{u_j}\ketbra{m}{q}\ketbra{u_j}{u_i}\ket{p}.
\end{align}
and 
\begin{align*}
    \tr[\rho_{pq}]\tr[\rho_{mn}] &= \bra{p}\tr_B[\rho]\ket{q}\bra{m}\tr_B[\rho]\ket{n}\\
    &=\sum_{i,j}\lambda_i^2\lambda_j^2\bra{p}\ketbra{u_i}{u_i}\ket{q}\bra{m}\ketbra{u_j}{u_j}\ket{n}.
\end{align*}
Thus,
\begin{align*}
    &\sum_{p,q,m,n}\tr[\rho_{qp}\rho_{nm}]\tr[\rho_{pq}]\tr[\rho_{mn}]\\
    =&\sum_{p,q,m,n}(\sum_{k,l}\lambda_k^2\lambda_l^2 \bra{n}\ketbra{u_k}{u_l}\ketbra{m}{q}\ketbra{u_k}{u_l}\ket{p})( \sum_{i,j}\lambda_i^2\lambda_j^2\bra{p}\ketbra{u_i}{u_i}\ket{q}\bra{m}\ketbra{u_j}{u_j}\ket{n})\\
    =&\sum_{p,q,m,n}\sum_{i,j,k,l}\lambda_i^2\lambda_j^2\lambda_k^2\lambda_l^2(\bra{n}\ketbra{u_k}{u_l}\ketbra{m}{q}\ketbra{u_k}{u_l}\ket{p}\bra{p}\ketbra{u_i}{u_i}\ket{q}\bra{m}\ketbra{u_j}{u_j}\ket{n})\\
    =&\sum_{q,m}\sum_{i,j,k,l}\lambda_i^2\lambda_j^2\lambda_k^2\lambda_l^2 \tr[\ketbra{u_k}{u_l}\ketbra{m}{q}\ketbra{u_k}{u_l}\ketbra{u_i}{u_i}\ket{q}\bra{m}\ketbra{u_j}{u_j}]\\
    =&\sum_{i,j,k,l}\lambda_i^2\lambda_j^2\lambda_k^2\lambda_l^2 \tr[\ketbra{u_k}{u_l}\ketbra{u_i}{u_i}]\tr[\ketbra{u_j}{u_j}\ketbra{u_k}{u_l}]\\
    =& \sum_{i}\lambda_i^8.
\end{align*}

Therefore, we have,
\begin{align*}
    &\sum_{p,q,m,n}(d_B\tr[\rho_{pq}\rho_{mn}]-\tr[\rho_{pq}]\tr[\rho_{mn}])\\
    =& (d_B^2+1)(\sum_i \lambda_i^4)^2-2d_B(\sum_i\lambda_i^8).
\end{align*}

So,
\begin{align}
    \var[\partial_\mu C_n] = \frac{8}{d_B(d_B^2-1)^2}\tr[H^2]((d_B^2+1)(\sum_i \lambda_i^4)^2-2d_B(\sum_i\lambda_i^8))
\end{align}
Since the $d_B$ is $2$, we can simplify the equation above as
\begin{align}
    \var[\partial_\mu C_n] & = \frac{4}{9}\tr[H^2](\lambda_1^8+\lambda_2^8 + 10\lambda_1^4\lambda_2^4) \\
    &=\frac{8}{9}(c_1^4 + c_2^4 + 10c_1^2c_2^2).
\end{align}
where the $c_1 = \lambda_1^2$, $c_2 = \lambda_2^2$ such that $c_1+c_2 = 1$, and $\tr[H^2] = d_B = 2$.

Therefore, we can simply get the range of the variance.
\begin{align}
    \frac{16}{27}\leq \var[\partial_\mu C_n] \leq \frac{8}{9}
\end{align}

%%%%%%%%%%%%%%%%%%%%%%%%%%%%%%%%%%%%%%%%%%%%%%%%%%%%%%%%%%%%%%%%%%%%%%%%%%%
% \section{Proof for Proposition~\ref{prop:main result on gradient of QSSM}}\label{appendix:name_of_appendix_A}
\subsection{Trainability of the middle step}
\begin{lemma}\label{lemma: gradient for fixed input}
For the target pure state $\rho_{ABC}$ on system $ABC$, suppose we start from a initial state $\hat{\sigma}$ such that $\tr_{BC}[\rho] = \tr_{BC}[\hat{\sigma}]$ and the output state is $\sigma$. If the cost function is
\begin{equation}
    C = \tr[(\tr_{C}[\rho]-\tr_{C}[\sigma])(\tr_{C}[\rho]-\tr_{C}[\sigma])]
\end{equation}
and the circuit is acting on system $BC$ while forming a local 4-design, then $\EE[\partial_\mu C] = 0$ and the variance of cost gradient scales as $\var[\partial_\mu C] \in \mathcal{O}(\frac{1}{d_B^3d_C})$, where $d_B, d_C$ denote the dimension of system $B$ and $C$ respectively.
\end{lemma}
Since $\tr_{BC}[\rho] = \tr_{BC}[\hat{\sigma}]$, there exist a fixed unitary $V$ such that
\begin{equation}
    \hat{\sigma}  = (I_A\otimes V_{BC})\rho(I_A\otimes V_{BC}^{\dagger}).
\end{equation}
Then
\begin{equation}
   \sigma  = (I\otimes UV)\rho(I\otimes V^{\dagger}U^{\dagger}).
\end{equation}
Then, the cost gradient becomes,
\begin{align*}
    \partial_\mu C &= 2\tr[\tr_C[\sigma]\partial_\mu \tr_{C}[\sigma]- 2\tr[\tr_C[\rho]\partial_\mu \tr_{C}[\sigma]]\\
    &= 2i\tr[\tr_C[(I\otimes U_{+}U_-V)\rho(I\otimes V^{\dagger}U_-^{\dagger}U_{+}^{\dagger})-\rho]\tr_C[(I\otimes U_{+}U_-V)\rho(I\otimes V^{\dagger}U_-^{\dagger}HU_{+}^{\dagger})\\ &- (I\otimes U_{+}HU_-V)\rho(I\otimes V^{\dagger}U_-^{\dagger}U_{+}^{\dagger})]]
\end{align*}
 We exploit the RTNI package~\cite{Fukuda2019} to calculate the mean of the cost gradient. It turns out that the mean of the cost gradient is zero.
 \begin{equation*}
     \mathbb{E}[\partial_\mu C] = 0
 \end{equation*}
 Then we consider the variance
 \begin{align*}
     \rm{Var}[\partial_\mu C] = - \mathbb{E}[(\partial_\mu C)^2]
 \end{align*}

With the RTNI package~\cite{fukuda2019rtni}, it turns out that the exact expression of the variance is dominant by 

 \begin{equation*}
 \operatorname{Var} [\partial_\mu C] \xrightarrow{d\rightarrow\infty} \frac{\tr[H^2]}{d_B^2(d_B^2d_C^2-1)} \cdot \left(
\diagram{
    % radius
    \def\r{0.5};
    % space
    \def\spacex{1.5};
    \def\spacey{1.5};
    % x-coordinate
    \def\xc{0};
    \def\xct{\xc+3*\r};
    \def\xrt{\xc+7*\r};
    \def\xlt{\xc-\r};
    % y-coordinate
    \def\yc{0};
    \def\yu{1*\spacey};
    \def\yuu{2*\spacey};
    \def\yuuu{3*\spacey};
    % coordinate of circle
    % point at center
    \coordinate (pc) at (\xct,\yc);
    % point upper
    \coordinate (pu) at (\xct,\yu);
    \coordinate (puu) at (\xct,\yuu);
    \coordinate (puuu) at (\xct,\yuuu);
 
    \coordinate (pxryuu) at (\xrt,\yuu);
    \coordinate (pxryu) at (\xrt,\yu);
    \coordinate (pxryuuu) at (\xrt+4*\r,\yuu);
 
    \coordinate (pxlyuu) at (\xlt, \yuu);
    \coordinate (pxlyu) at (\xlt, \yu);
    
    % control Bezier curves
    \draw (\xlt-\r,\yuuu-0.5*\r) .. controls (\xlt-2*\r,\yuuu-0.5*\r) and (\xlt-2*\r,\yuu+0.5*\r) .. (\xlt-\r,\yuu+0.5*\r);
          \draw (\xct+\r,\yuuu-0.5*\r) .. controls (\xct+2*\r,\yuuu-0.5*\r) and (\xct+2*\r,\yuu+0.5*\r) .. (\xct+\r,\yuu+0.5*\r);

        \draw (\xrt-\r,\yuuu-0.5*\r) .. controls (\xrt-2*\r,\yuuu-0.5*\r) and (\xrt-2*\r,\yuu+0.5*\r) .. (\xrt-\r,\yuu+0.5*\r);
          \draw (\xrt+5*\r,\yuuu-0.5*\r) .. controls (\xrt+6*\r,\yuuu-0.5*\r) and (\xrt+6*\r,\yuu+0.5*\r) .. (\xrt+5*\r,\yuu+0.5*\r);
          
    \draw (\xlt-\r,\yuu-1.5*\r) .. controls (\xlt-1.5*\r,\yuu-1.5*\r) and (\xlt-1.5*\r,\yu) .. (\xlt-\r,\yu);
        \draw (\xrt+3*\r,\yuu-1.5*\r) .. controls (\xrt+2.5*\r,\yuu-1.5*\r) and (\xrt+2.5*\r,\yu) .. (\xrt+3*\r,\yu);
        
        \draw (\xlt+\r,\yuu-1.5*\r) .. controls (\xlt+1.5*\r,\yuu-1.5*\r) and (\xlt+1.5*\r,\yu) .. (\xlt+\r,\yu);
        \draw (\xrt+5*\r,\yuu-1.5*\r) .. controls (\xrt+5.5*\r,\yuu-1.5*\r) and (\xrt+5.5*\r,\yu) .. (\xrt+5*\r,\yu);

         \draw (\xct-\r,\yuu-1.5*\r) .. controls (\xct-1.5*\r,\yuu-1.5*\r) and (\xct-1.5*\r,\yuu-2.5*\r) .. (\xct-\r,\yuu-2.5*\r);
         \draw (\xrt+\r,\yuu-1.5*\r) .. controls (\xrt+1.5*\r,\yuu-1.5*\r) and (\xrt+1.5*\r,\yuu-2.5*\r) .. (\xrt+\r,\yuu-2.5*\r);

         \draw (\xct-\r,\yuu-0.5*\r) .. controls (\xct-2*\r,\yuu-0.5*\r) and (\xct-2*\r,\yuu-3.5*\r) .. (\xct-\r,\yuu-3.5*\r);
         \draw (\xrt+\r,\yuu-0.5*\r) .. controls (\xrt+2*\r,\yuu-0.5*\r) and (\xrt+2*\r,\yuu-3.5*\r) .. (\xrt+\r,\yuu-3.5*\r);

          \draw (\xlt-\r,\yuu-0.5*\r) .. controls (\xlt-2*\r,\yuu-0.5*\r) and (\xlt-2*\r,\yuu-4.5*\r) .. (\xlt-\r,\yuu-4.5*\r);
         \draw (\xrt+5*\r,\yuu-0.5*\r) .. controls (\xrt+6*\r,\yuu-0.5*\r) and (\xrt+6*\r,\yuu-4.5*\r) .. (\xrt+5*\r,\yuu-4.5*\r);

          \draw (\xlt+\r,\yuu-0.5*\r) .. controls (\xlt+2*\r,\yuu-1.5*\r) and (\xlt+2*\r,\yuu-4*\r) .. (\xlt+3*\r,\yuu-4*\r);

          \draw (\xrt+3*\r,\yuu-0.5*\r) .. controls (\xrt+2*\r,\yuu-1.5*\r) and (\xrt+2*\r,\yuu-4*\r) .. (\xrt+\r,\yuu-4*\r);
    
    % horizontal line
        \draw (\xc-2*\r,\yuuu-0.5*\r) -- (\xc+4*\r,\yuuu-0.5*\r);
        \draw (\xc-2*\r,\yuu+0.5*\r) -- (\xc+4*\r,\yuu+0.5*\r);

         \draw (\xc-2*\r,\yu) -- (\xc,\yu);
        
         \draw (\xrt-\r,\yuuu-0.5*\r) -- (\xrt+5*\r,\yuuu-0.5*\r);
        \draw (\xrt-\r,\yuu+0.5*\r) -- (\xrt+5*\r,\yuu+0.5*\r);

        \draw (\xrt+3*\r,\yu) -- (\xrt+5*\r,\yu);

        \draw (\xct,\yuu-0.5*\r) -- (\xrt,\yuu-0.5*\r);
        \draw (\xct,\yuu-1.5*\r) -- (\xrt,\yuu-1.5*\r);
        
        \draw (\xct-\r,\yuu-4*\r) -- (\xrt+\r,\yuu-4*\r);

        \draw (\xct-\r,\yuu-2.5*\r) -- (\xrt+\r,\yuu-2.5*\r);
        \draw (\xct-\r,\yuu-3.5*\r)-- (\xrt+\r,\yuu-3.5*\r);

        \draw (\xc-2*\r,\yuu-4.5*\r) -- (\xrt+5*\r,\yuu-4.5*\r);

    % tensor square
    \draw[ten, shift=(puu)] (-\r,-2*\r) rectangle (\r,\r);
    \node at (\xct,\yuu) {\scriptsize $\rho$};
    \draw[ten, shift=(pxryuu)] (-\r,-2*\r) rectangle (\r,\r);
    \node at (pxryuu) {\scriptsize $\rho$};
    \draw[ten, shift=(pxlyuu)] (-\r,-2*\r) rectangle (\r,\r);
    \node at (pxlyuu) {\scriptsize $\rho$};  
        \draw[ten, shift=(pxryuuu)] (-\r,-2*\r) rectangle (\r,\r);
    \node at (pxryuuu) {\scriptsize $\rho$};

}\right)
\end{equation*}
We know that $\tr[H^2] = d_Bd_C$, thus we have
\begin{equation*}
\var[\partial_\mu C]\in \mathcal{O}(\frac{g(\rho)}{d_B^3d_C}),
\end{equation*}
where $g(\rho)$ denotes the dominant factor from the tensor product illustrated above. Finally, we can conclude the following Proposition,

\begin{proposition}\label{proposition: gradient with layer width}
    For the $k$-th learning step ($k\leq n$) in QSSM, the mean of cost gradient is 0, and the variance of cost gradient scales as  $\var[\partial_\mu C_k] \in \mathcal{O}(2^{-n_k})$, where $n_k$ is the circuit width of $k$-th learning step.
\end{proposition}

Suppose the target state is $\rho$ and the input state for $k$-the learning step is $\hat{\sigma}$. 
We assume system $A$ denotes the first $k-1$ qubits, system $B$ denotes the $k$-th qubit and system $C$ denotes the $(k+1)$-th qubit to the $(k+n_k-1)$-th qubit. With the definition of $n_k$ claimed in the text, there exists a purification $\hat{\rho}_{ABC}$ of $\rho_A$ on system $ABC$. According to lemma~\ref{lemma: gradient for fixed input}, we can easily know that 
\begin{equation*}
    \var[\partial_\mu C_k] \in \mathcal{O}(\frac{1}{2^{n_k+2}}) = \mathcal{O}(2^{-n_k})
\end{equation*}

\begin{proposition}\label{prop:main result on gradient of QSSM in app}[Trainability]
Given the state learning algorithm stated in Proposition~\ref{proposition: Truncation}, for an $n$-qubit pure target state $\rho$ represented by $n$ ordered quantum registers $q_1, q_2, \cdots, q_n$ with a rank sequence $\cR_\rho = \{r_1, r_2, \cdots r_{n-1},r_n\}$, if one of the $U_{\pm}^{(k)}$ in the $k$-th scattering layer $U_k$ forms at least local unitary $4$-design, the expectation and the variance of $C_k$ with respect to $\theta_\mu$ can be upper bounded by,
\begin{equation*}
    \EE[\partial_{\mu} C_k] = 0; \quad
    \var[\partial_{\mu} C_k] \in \cO\left(\frac{g(\rho_k)}{r_k}\right),
\end{equation*}
where the expectation is computed regarding the Haar measure and the factor $g(\rho_k)$ scales polynomially in $\tr[\rho_k^2]$ known as the purity of $\rho_k$.
\end{proposition}

% \begin{proposition}[Main Prop.~\ref{prop:main result on gradient of QSSM}]
% For a $n$-qubit target state $\rho$ with fixed-order representation, we suppose its rank sequence is $\cR_\rho = \{r_1, r_2, \cdots r_{n-1},r_n\}$. Then, for learning the target state $\rho$ with QSSM, if the circuit used for each step is sufficiently random that forms a local 4-design, the expectation gradient for the $k$-th step $\E[\partial_{\mu} C_k] = 0$ and the variance of the cost gradient scales with $r_k$ as,
% \begin{equation}
%     \var[\partial_{\mu} C_k] \in \mathcal{O}(\frac{1}{r_k}).
% \end{equation}
% \end{proposition}

Since we know that $2^{n_k-1}\leq r_k \leq 2^{n_k}$, thus according to 
% proposition~\ref{proposition: last step gradient} and 
Proposition~\ref{proposition: gradient with layer width}, we can get the proof. Notice that the factor $g(\rho_k)$ scales polynomially in $\tr[\rho_k^2]$ due to the Cauchy-Schwartz inequality of density matrices. We then finish the proof of the Proposition.

\section{Analytic evaluation of cost function and gradient}\label{appendix:analytic_cost}
\begin{figure}[!h]
    \centering
    \includegraphics[width=0.5\linewidth]{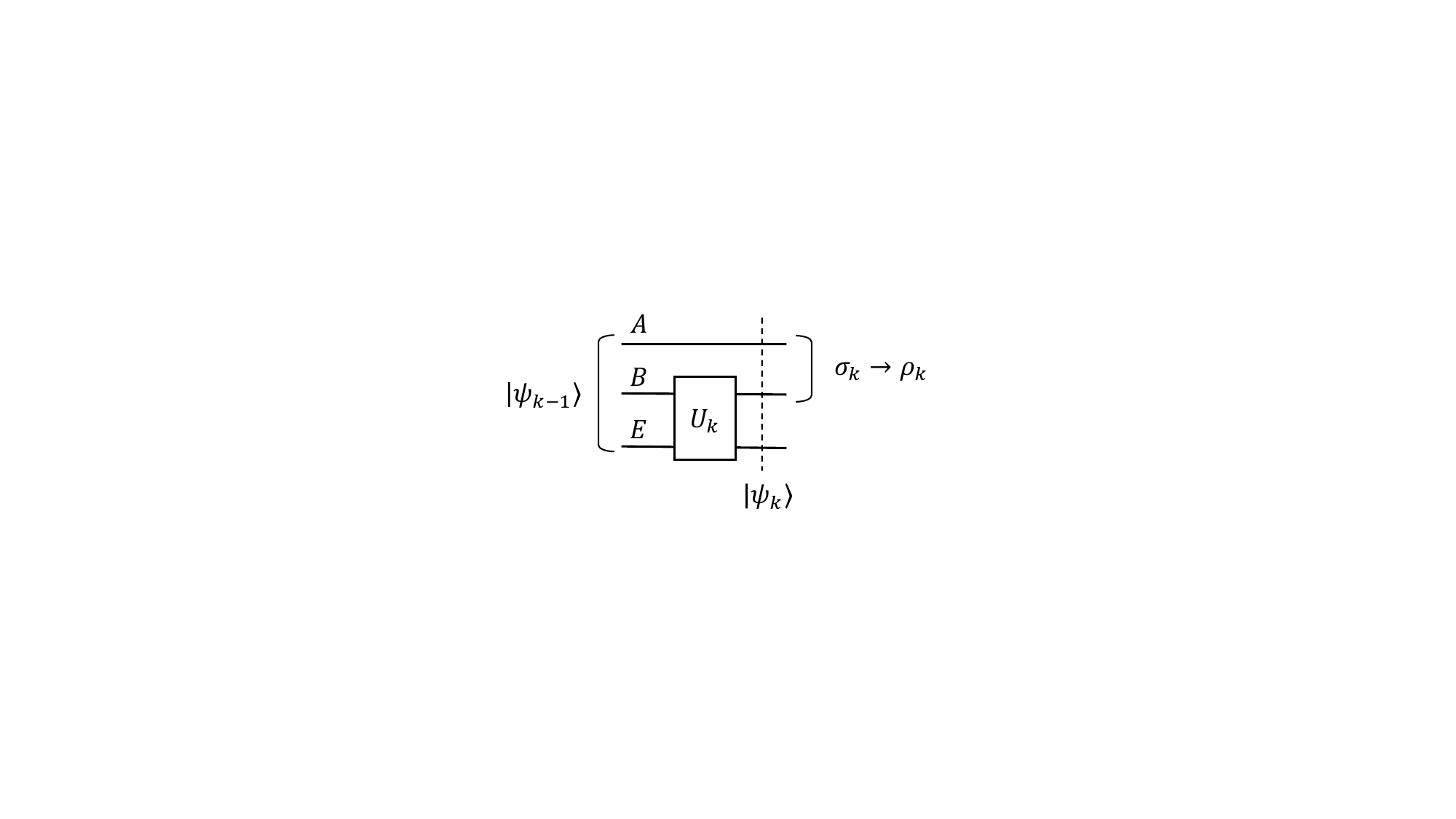}
    \caption{The $k$-th learning step layer. Based on adaptive learning processes, the previously learnt state $\ket{\psi_{k-1}}$ on system $ABE$ must be pure where $E$ is the additional system acted by the $k$-th step layer $U_k$. Under perfect learning situation, we have $\sigma_{k-1} = \tr_{BE}(\psi_{k-1}) = \rho_{k-1}$.}
    \label{Fig: grad_fig1}
\end{figure}

In this appendix, we provide a detailed analysis of the analytic gradient of our cost function $C_k$~\eqref{Eq: cost_func}. We take the 2-norm squared cost function as our objective. At the $k$-th learning step, analyzing the exact form of $\partial_\mu C_k$ is necessary for further designing the training strategy of QSSM. Recalling the expression of $C_k$, we could derive the derivative form with respect to the parameter $\theta_\mu = \theta_k^\mu$. From here, we have concentrated on the $k$-th step and for convenience, we will omit the subscript $k$ of the parameter in the following sections. The partial derivative of $C_k$ with respect to $\theta_\mu$ is then expressed as,
\begin{equation}\label{Eq: grad_expression1}
    \partial_\mu C_k = 2\tr(2\sigma_k \partial_\mu(\sigma_k)) - 2\tr(\rho_k \partial_\mu(\sigma_k)),
\end{equation}
where $\sigma_k = \sigma_k(\bm{\theta})$ which is constructed via paramterized circuit $U_k(\bm{\theta})$, and $\rho_k$ is the $k$-th step reduced target. In a practical sense, our $U_k$ is composed of the quantum gates satisfying the parameter-shift rule and $U_k = U_l e^{-i\frac{\theta_\mu}{2}\Omega_\mu} U_r = \Tilde{U}_l U_r$, where $\Omega_\mu^2 = I$. The $k$-th scattering layer has been shown in Fig.~\ref{Fig: grad_fig1}. Then the following lemma holds,
\begin{lemma}\label{lemma: exact form of partial mu cost}
    The $k$-th step cost function $C_k$ has the partial derivative form (w.r.t. $\theta_\mu$ and evaluated at $\bm{\theta} = \bm{\theta}^*$),
    \begin{equation*}
        \partial_\mu C_k^* = \left\langle\Delta_k^* \otimes \frac{I_E}{d_E}\right\rangle_{\theta_\mu + \frac{\pi}{2}} - \left\langle\Delta_k^*\otimes \frac{I_E}{d_E}\right\rangle_{\theta_\mu - \frac{\pi}{2}}
    \end{equation*}
    where $\Delta_k = \sigma_k - \rho_k$ with $*$ indicating the state difference evaluated at $\bm{\theta}^*$. The other symbols all match the settings in Fig.~\ref{Fig: grad_fig1}. 
\end{lemma}
By observing $\sigma_{k} = \tr_E((I_A\otimes U_k)P_{\psi_{k-1}}(I_A\otimes U_k^\dag))$, where $P_{\psi_{k-1}} = \ketbra{\psi_{k-1}}{\psi_{k-1}}$, we could compute the expression of $\partial_\mu \sigma_k$ based on the linearity of derivative operation,
\begin{equation*}
    \partial_\mu \sigma_k = \tr_E((I_A\otimes \partial_\mu(U_k))P_{\psi_{k-1}}(I_A\otimes U_k^\dag)) + \tr_E((I_A\otimes U_k)P_{\psi_{k-1}}(I_A\otimes \partial_\mu(U_k^\dag))).
\end{equation*}
Recalling the expression of $\partial_\mu (U_k)$ and $\partial_\mu (U_k^\dag)$, we have,
\begin{equation*}
\begin{aligned}
    \partial_\mu \sigma_k &= -\frac{i}{2}\tr_E((I_A\otimes \Tilde{U}_l)[(I_A\otimes\Omega_\mu), (I_A\otimes U_r)P_{\psi_{k-1}}(I_A\otimes U_r^\dag)](I_A\otimes \Tilde{U}^\dag_l)) \\ 
    &= -\frac{i}{2}\tr_E(\Tilde{U}_l[\Omega_\mu, U_rP_{\psi_{k-1}}U_r^\dag] \Tilde{U}^\dag_l)
\end{aligned}, 
\end{equation*}
where we have abbreviated the `$I_A\otimes$' correspondence for simplicity, which the subsystem $A$ would never join the optimizations during the $k$-th step. Since $U_\mu(\theta_\mu) = e^{-i\frac{\theta_\mu}{2}\Omega_\mu}$ satisfies the parameter-shift rule. we could use the gate identity,
\begin{equation*}
    i[\Omega_\mu, M] = U_{\mu}\left(-\frac{\pi}{2}\right) M U_{\mu}^\dag\left(-\frac{\pi}{2}\right) - U_{\mu}\left(\frac{\pi}{2}\right) M U_{\mu}^\dag\left(\frac{\pi}{2}\right)
\end{equation*}
for any linear operator $M$, and then derive the exact value of $\partial_\mu \sigma_k^*$ at $\bm{\theta} = \bm{\theta}^*$ as,
\begin{equation*}
    \partial_\mu(\sigma_k^*) = \frac{1}{2}\tr_E\left(U_k(\theta_\mu^* + \frac{\pi}{2})P_{\psi_{k-1}}U^\dag_k(\theta_\mu^* + \frac{\pi}{2}) - U_k(\theta_\mu^* - \frac{\pi}{2})P_{\psi_{k-1}}U^\dag_k(\theta_\mu^* - \frac{\pi}{2})\right).
\end{equation*}
Here $\partial_\mu(\sigma_k^*) = \partial_\mu(\sigma_k)|_{\bm{\theta} = \bm{\theta}^*}$, and circuit $U_k(\theta_\mu^* + \alpha)$ intakes $\bm{\theta^*}$ and modifies the parameter $\theta_\mu^*$ to $\theta_\mu^*+\frac{\pi}{2}$. Now, recalling the fact that,
\begin{equation*}
    \tr(\tr_B(\rho_{AB})\sigma_A) = \tr\left(\rho_{AB}(\sigma_A \otimes \frac{I_B}{d_B})\right),
\end{equation*}
we have,
\begin{equation*}
\begin{aligned}
    \tr(\rho_k\partial_\mu(\sigma_k^*)) &= \left\langle\rho_k\otimes \frac{I_E}{d_E}\right\rangle_{\theta_\mu^* + \frac{\pi}{2}} - \left\langle\rho_k\otimes \frac{I_E}{d_E}\right\rangle_{\theta_\mu^* - \frac{\pi}{2}}\\
    \tr(\sigma^*_k\partial_\mu(\sigma_k^*)), &= \left\langle\sigma^*_k\otimes \frac{I_E}{d_E}\right\rangle_{\theta_\mu^* + \frac{\pi}{2}} - \left\langle\sigma^*_k\otimes \frac{I_E}{d_E}\right\rangle_{\theta_\mu^* - \frac{\pi}{2}},
\end{aligned}
\end{equation*}
where $\langle M \rangle_\theta = \bra{\psi_k(\theta)} M \ket{\psi_k(\theta)}$ and $\ket{\psi_k(\theta)}$ is derived by applying $U_k(\theta)$ on $\ket{\psi_{k-1}}$. Combining the above calculations to obtain the desired result in lemma~\ref{lemma: exact form of partial mu cost} taking $\Delta^* = \sigma_k(\bm{\theta}^*) - \rho_k$. Finally, by taking the actual dimensional factors, we could derive the analytic form of the partial derivative as shown in Sec.~\ref{subsec: cost and gradient}.
\end{document}